\documentclass[journal]{IEEEtran}
\IEEEoverridecommandlockouts
\usepackage{cite}
\usepackage{amsmath,amssymb,amsfonts}
\usepackage{algorithmic}
\usepackage{graphicx}
\usepackage{textcomp}
\usepackage[table]{xcolor}
\usepackage{authblk}
\usepackage{url}
\usepackage{verbatim}

\usepackage[nolist]{acronym}
\begin{acronym}
    \acro{DoD}[DoD]{degree of decentralization}
    \acro{HHI}[HHI]{Herfindahl–Hirschman Index}
    \acro{JSON}[JSON]{JavaScript Object Notation}
    \acro{PoS}[PoS]{proof-of-stake}
    \acro{PoW}[PoW]{proof-of-work}
    \acro{P2P}[P2P]{peer-to-peer}
    \acro{SBR}[SBR]{stale block rate}
    \acro{TPS}[TPS]{transactions per second}
    \acro{BFT}[BFT]{Byzantine fault-tolerance}
    \acro{PBFT}[PBFT]{Practical Byzantine fault-tolerance}
    \acro{MTBF}[MTBF]{Mean Time Between Failures}
    \acro{MTTR}[MTTR]{Mean Time To Repair}
    \acro{MEV}[MEV]{maximum extractable value}
\end{acronym}
\usepackage{tabularx}
\usepackage{multirow}
\usepackage{tabularray}
\usepackage{rotating}
\usepackage{lscape}
\usepackage[hidelinks]{hyperref}
\usepackage{amsmath}
\usepackage{placeins}
\usepackage{float}
\usepackage{microtype}
\graphicspath{ {sections/figures} }
\newcolumntype{Y}{>{\centering\arraybackslash}X}
\usepackage{fancyhdr}
\usepackage{booktabs}
\usepackage{makecell}

\usepackage{balance}

\definecolor{fullgreen}{HTML}{DDEBC8}
\definecolor{partorange}{HTML}{F8C98B}
\definecolor{nonered}{HTML}{E58E88}
\newcommand{\full}{\cellcolor{fullgreen}{+}}
\newcommand{\partly}{\cellcolor{partorange}{$\pm$}}
\newcommand{\none}{\cellcolor{nonered}{--}}

\newcommand{\BottomHeader}[2]{%
  \begin{minipage}[b][#1][b]{\linewidth}
  \centering\bfseries #2
  \end{minipage}
}

\newcommand{\constructDefinition}[2]{
\vspace{10pt} \noindent \textbf {#1}: \textit{#2} }

\def\BibTeX{{\rm B\kern-.05em{\sc i\kern-.025em b}\kern-.08em
		T\kern-.1667em\lower.7ex\hbox{E}\kern-.125emX}}
\begin{document}

     \title{From Concept to Measurement: A Survey of How the Blockchain Trilemma Can Be Analyzed}

    \author[1]{Mansur Masama Aliyu}
    \author[2]{Niclas Kannengießer}
    \author[3]{Ali Sunyaev}
    \affil[1]{Karlsruhe Institute of Technology, Karlsruhe, Germany \authorcr Email: {\tt \{mansur.masama\}@partner.kit.edu}\vspace{1.5ex}}
    \affil[2]{Karlsruhe Institute of Technology, Karlsruhe, Germany \authorcr Email: {\tt \{niclas.kannengiesser\}@kit.edu}\vspace{1.5ex}}
    \affil[3]{Technical University of Munich, Campus Heilbronn, Germany \authorcr Email: {\tt \{ali.sunyaev\}@tum.de}\vspace{1.5ex}}

	\maketitle

	\begin{abstract}
The blockchain trilemma provides an analytical lens for examining trade-offs among \ac{DoD}, scalability, and security in blockchain systems. However, different aspects of these dimensions are measured in different ways, making it difficult to justify metric selection, compare results across studies, and interpret what a metric value means for a trilemma dimension. Through a review of blockchain-trilemma-related literature, we synthesize 12 measurable aspects and 15 metrics for analyzing \ac{DoD}, scalability, and system-level security in core blockchain systems. We clarify which aspects of the blockchain trilemma dimensions these metrics quantify, when they are applicable, and how their interpretation depends on blockchain system designs and analysis approaches. These results help researchers and practitioners select suitable metrics, interpret metric values, and compare blockchain systems with respect to trade-offs among \ac{DoD}, scalability, and security.
\end{abstract}

	\begin{IEEEkeywords}
    Benchmark, blockchain technology, trade-offs, quality requirements.
\end{IEEEkeywords}

    \section{Introduction}
\label{sec:introduction}

Blockchain system development requires design decisions that often affect multiple system properties.
For example, increasing transaction throughput by assigning consensus-critical tasks to fewer validating nodes per decision can improve scalability but reduce \ac{DoD} when influence on consensus becomes concentrated among fewer validating nodes.
For targeted blockchain system design, software architects need to reason about such coupled effects across \ac{DoD}, scalability, and security.
The blockchain trilemma provides an analytical lens for such reasoning by framing recurring tensions among these dimensions and by expressing the view that they are difficult to maximize simultaneously~\cite{xiao2020survey, nakai2024formulation, werth2023review, monte2020scaling, quattrocchi2024blockchain}.

The blockchain trilemma helps formulate expectations about how design decisions may shift the balance among these dimensions.
Yet the trilemma is abstract and does not specify which system properties should be observed or which metrics can provide empirical evidence for these expectations.
Empirically testing such trilemma-based expectations therefore requires translating abstract reasoning about \ac{DoD}, scalability, and security into observable properties of blockchain system designs.

This translation is difficult because a trilemma dimension can be represented by different constructs, and each construct can be measured in different ways.
For example, if a design assigns consensus-critical tasks to fewer validating nodes, trilemma-based reasoning suggests a reduction in \ac{DoD}.
To test this expectation, \ac{DoD} must be represented by constructs that capture specific aspects of influence on consensus.
Block-proposal randomness can capture one such aspect when block proposal reflects consensus influence, and Shannon entropy can operationalize the resulting distribution~\cite{wu2019information, yan2025data}.
Wealth distribution, operationalized through the Gini coefficient~\cite{juodis2024overview, ovezik2025sok}, provides weaker evidence for the same expectation when token holdings do not shape influence on consensus.
Conversely, block-proposal randomness provides incomplete evidence when block proposal is separated from other consensus-critical tasks.
Thus, a metric value may be correct while still supporting only limited conclusions about the trilemma dimension that motivated the measurement.
For meaningful empirical analysis of blockchain systems, software architects therefore need to understand which aspects of a trilemma dimension a construct captures, how metrics operationalize that construct, and under which applicability conditions the measured values support conclusions about the dimension.

Existing analysis approaches and metric definitions only partially support such understanding.
For example, \textit{BBSF}~\cite{ren2023bbsf}, \textit{BLOCKBENCH}~\cite{dinh2017blockbench}, \textit{Diablo}~\cite{gramoli2023diablo}, and simulators such as \textit{SimBlock}~\cite{aoki2019simblock, nakai2024formulation} show how selected properties of blockchain systems can be measured.
Other studies define metrics for constructs, such as block-proposal randomness~\cite{chemaya2025dataset, wu2019information, yan2025data} and wealth distribution~\cite{juodis2024overview, nakai2024formulation, quattrocchi2024blockchain, ovezik2025sok}.
Yet these contributions do not consistently explain which construct a metric operationalizes, when it applies and how metric values should be interpreted.
This makes construct and metric selection difficult to justify in trilemma-oriented empirical analyses.
To support more consistent and defensible construct and metric selection, we pose the following research question:
\emph{Which constructs and associated metrics are suitable for analyzing blockchain system designs with respect to the blockchain trilemma dimensions?}

We conducted a concept-centric systematic literature search to identify publications that propose constructs and metrics for analyzing \ac{DoD}, scalability, and security~\cite{webster2002analyzing}.
We then analyzed the collected literature using framework-guided deductive--inductive thematic synthesis~\cite{braun2006using, thomas2008methods}.
The blockchain trilemma served as an analytical lens: its three dimensions provided deductive top-level categories, while lower-level constructs and metrics were inductively synthesized from the reviewed literature.
To improve conceptual coverage, we complemented the systematic literature search with purposive sampling of information-rich publications~\cite{palinkas2015purposeful}.
Based on this analysis, we synthesized constructs, metrics, applicability conditions, input parameters, limitations, and analysis approaches used in prior literature.

This work contributes to trilemma-oriented empirical analysis and benchmarking of blockchain systems in three ways.
\textit{First}, we synthesize construct--metric combinations for analyzing blockchain systems with respect to \ac{DoD}, scalability, and security. This synthesis distinguishes the aspect of a trilemma dimension being analyzed from the metric used to operationalize it.
\textit{Second}, we characterize construct--metric combinations by their applicability conditions and limitations. This characterization clarifies under which blockchain system designs measured values support conclusions about a trilemma dimension.
\textit{Third}, we identify the input parameters required by the metrics and distinguish design from environmental input parameters. This supports data collection, benchmark design, and metric interpretation.
In addition, we relate construct--metric combinations to analysis approaches used in prior literature, showing how existing analytical, simulation-based, and benchmarking approaches operationalize trilemma-related constructs.

The remainder of this work is structured as follows.
Section~\ref{sec:back-and-rw} introduces conceptual foundations of blockchain technology, including the system boundary used in this work, the blockchain trilemma as an analytical lens, and existing conceptual and empirical attempts to operationalize its dimensions.
Section~\ref{sec:research-approach} describes our literature search, thematic synthesis, and external feedback on construct descriptions, metric specifications, and mappings.
Section~\ref{sec:results} presents the identified constructs and metrics associated with the blockchain trilemma dimensions, and reports the applicability conditions of construct--metric combinations.
Section~\ref{sec:discussion} discusses our findings, contributions to practice and research, limitations, and open research directions.
Finally, Section~\ref{sec:conclusion} concludes with key takeaways.
    \section{Conceptual Foundations and Related Research}
\label{sec:back-and-rw}

This section introduces the conceptual foundations used in this work.
First, we define the blockchain system boundary.
Second, we summarize system-design properties that shape how \ac{DoD}, scalability, and security can be analyzed.
Third, we specify the blockchain trilemma dimensions and their interrelationships.
Finally, we position this work relative to prior research on trilemma-oriented measurement of blockchain systems.

\subsection{Blockchain Technology and System Model}
\label{sec:foundations-of-bc-technology}

Blockchain systems are replicated ledger systems in which consensus or ordering protocols establish a canonical order of state-changing transactions.
A \emph{transaction originator} is an external entity, such as a client application or end-user wallet, that creates and submits transactions to the blockchain system.
Transaction originators do not participate in consensus or maintain the ledger.
They interact with the system through entry and propagation interfaces of validating nodes, such as JavaScript Object Notation--Remote Procedure Call (JSON--RPC), Google--Remote Procedure Call (gRPC), or Quick User Datagram Protocol Internet Connections (QUIC).

In this work, we focus on the \emph{core blockchain system}, that is, the consensus-critical infrastructure responsible for transaction propagation, ordering, validation, execution, and replication.
The core blockchain system consists of \emph{validating nodes}.
We use \emph{validating node} as a generic umbrella term rather than as a protocol-specific role name.
Validating nodes contribute to one or more consensus-critical tasks, including transaction propagation, block proposal, transaction ordering, or ledger replication.
Not all validating nodes perform every task: some participate only in ordering, whereas others also validate blocks, execute transactions, update ledger state, and store replicated data.

Application-layer artifacts, including smart contracts deployed to blockchain systems and their vulnerabilities, are outside the scope of this work.
Accordingly, the security dimension in this work refers to system-level security.

\begin{figure*}[!t]
    \centering
    \includegraphics[width=0.75\textwidth]{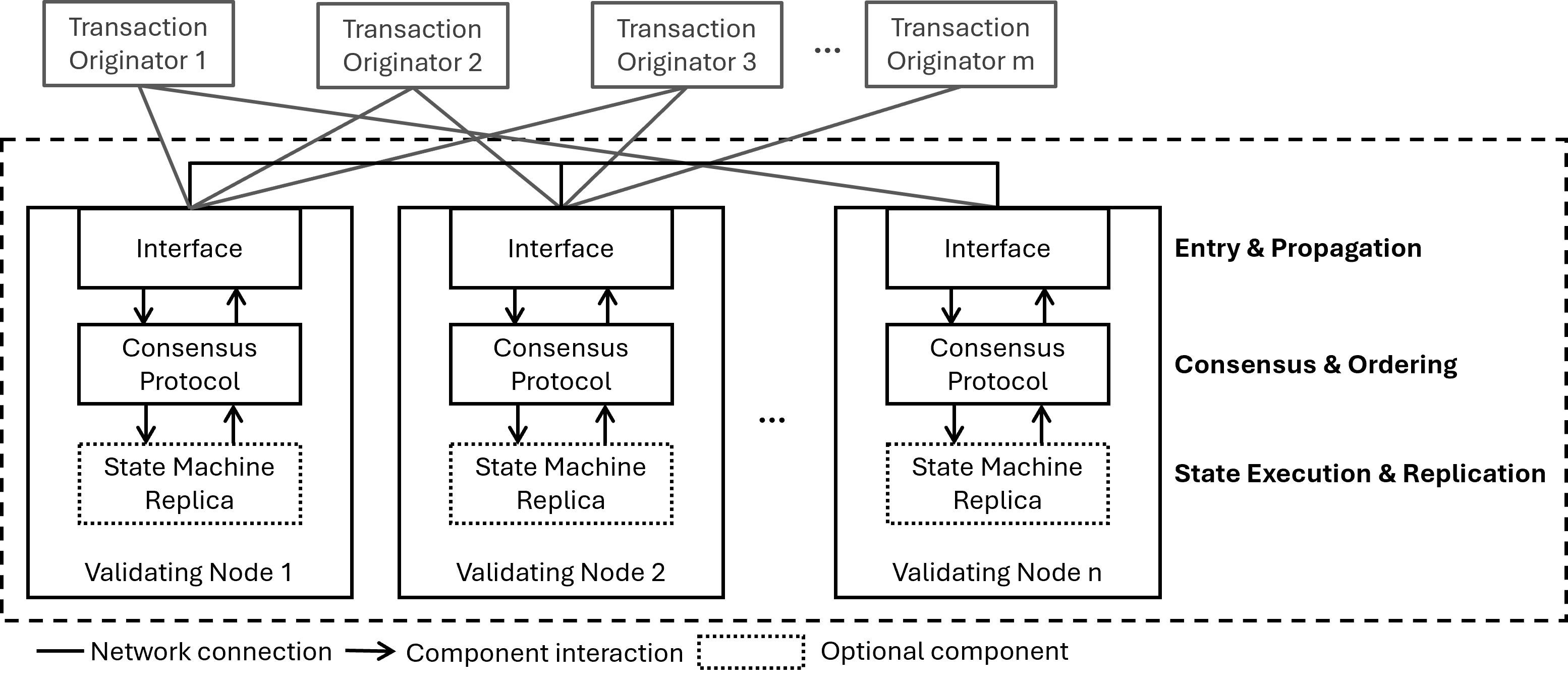}
    \caption{Layered blockchain system model adapted from Leinweber et al.~\cite{leinweber2023leveraging}. 
    The dashed border delineates the \emph{core blockchain system} composed of \emph{validating nodes}; 
    \emph{transaction originators} (clients/applications, end-user wallet, or service) submit transactions from outside this boundary. 
    Validating nodes implement one or more functional layers: 
    \emph{Entry \& Propagation} (interfaces and dissemination), \emph{Consensus \& Ordering} (canonical ordering/finality), and 
    \emph{State Execution \& Replication} (deterministic validation/execution and ledger updates). 
    Not all validating nodes implement every layer; for example, Hyperledger Fabric \emph{orderers} participate only in Consensus \& Ordering and do not execute transactions or store the application ledger, whereas Fabric \emph{peers} and validators in permissionless systems implement the replicated state machine.
    Application-layer artifacts, including smart contract vulnerabilities, are outside the scope of this work.}
    \label{fig:blockchain-system-arch}
\end{figure*}

To describe heterogeneous blockchain systems uniformly, we distinguish three functional layers of the blockchain system.
\emph{Entry and Propagation} covers transaction-submission interfaces and dissemination.
\emph{Consensus and Ordering} covers protocol mechanisms that establish canonical ordering and finality.
\emph{State Execution and Replication} covers deterministic validation, transaction execution, ledger-state updates, and data replication.
Figure~\ref{fig:blockchain-system-arch} illustrates this layered model.
Table~\ref{tab:systemmodel-instance} shows how selected blockchain systems instantiate the layers.

\begin{table*}[!t]
\centering
\renewcommand{\arraystretch}{1.3}
\footnotesize
\caption{Instantiation of the layered model for selected systems. 
Each row maps system-specific components to 
(i) \emph{Entry \& Propagation}, 
(ii) \emph{Consensus \& Ordering}, and 
(iii) \emph{State Execution \& Replication}.
Only validating nodes that execute transactions and update the ledger implement the replicated state machine (e.g., peers, full nodes, validators); some validating nodes (e.g., Fabric orderers) do not. BFT = Byzantine Fault Tolerance, FFG = Friendly Finality Gadget, gRPC = Google–Remote Procedure Call, JSON–RPC = JavaScript Object Notation–Remote Procedure Call, LMD-GHOST = Latest Message Driven – Greediest Heaviest Observed SubTree, QUIC = Quick User Datagram Protocol Internet Connections, UTXO = Unspent Transaction Output.}
\label{tab:systemmodel-instance}
\resizebox{\linewidth}{!}{%
\begin{tblr}{
    rows = {valign=m},
    width = \textwidth,
    rows = {valign=m},
    column{1} = {2.6cm,l},
    column{2} = {3.6cm,l},
    column{3} = {co=1,l},
    column{4} = {co=1,l},
    colsep = 4pt,
    rowsep = 2pt,
    row{1} = {font=\bfseries,halign=c},
    hline{1,Z} = {1-4}{0.8pt},
    hline{2} = {1-4}{0.6pt},
    hline{3-Y} = {1-4}{0.25pt},
    vline{1,Z} = {1-Z}{0.8pt},
    vline{2-4} = {1-Z}{0.25pt}
}
\textbf{System} 
& \textbf{Entry \& Propagation} 
& \textbf{Consensus \& Ordering} 
& \textbf{State Execution \& Replication} 
\\

Bitcoin~\cite{bitcoin2020arch, bitcoin2025arch, nakamoto2008bitcoin} 
& JSON-RPC, mempool, gossip 
& \ac{PoW} + longest-work fork choice; \newline probabilistic finality 
& Full nodes validate blocks and update the UTXO set (miners typically also run full nodes)
\\

Ethereum~\cite{ethereum2025arch} 
& JSON-RPC, gossip 
& Gasper (combining LMD-GHOST fork choice with Casper-FFG epoch finality)
& Validators (consensus and execution clients) \newline validate transactions, execute state transitions, and update the ledger state
\\

Hyperledger \newline Fabric~\cite{hyperledger2025arch, guggenberger2022depth} 
& Client$\to$peer endorsement gRPC; client$\to$orderer submit 
& Ordering service (Raft/BFT) establishes total order; deterministic finality 
& Peers endorse transactions, validate ordered transactions, \newline and commit ledger/world state; orderers do not store the application ledger 
\\

Polygon \ac{PoS}~\cite{polygon2025arch}
& RPC/mempool, gossip 
& Bor block production + Heimdall checkpointing \newline to the Ethereum system
& Bor validators execute and commit locally; \newline Heimdall coordinates checkpointing
\\

Solana~\cite{solana2019update, solana2023interface}
& RPC, QUIC, Turbine
& PoH leader schedule + Tower BFT voting; fast confirmations 
& Validators replay/execute and commit ledger (shreds$\to$blocks) 
\\

\end{tblr}
}
\end{table*}

The selected systems illustrate how the system model applies to different designs within the scope of this work.
Bitcoin, Ethereum, Hyperledger Fabric, Polygon~\ac{PoS}, and Solana systems assign block production, ordering, validation, and replication to different role structures and provide finality through different protocol mechanisms~\cite{nakamoto2008bitcoin, sakurai2024tie, ethereum2025arch, hyperledger2025arch, guggenberger2022depth, polygon2025arch, solana2019update, solana2023interface}.
For example, Hyperledger Fabric separates orderers from peers, whereas Ethereum validators participate in proposal and attestation while relying on paired consensus and execution clients.

\subsection{Interrelationships Between Consensus Protocols and Key Properties of Blockchain Systems}
\label{sec}

Consensus protocols are closely intertwined with key properties of blockchain systems, including permission models, finality models, fault models, and performance.
These properties are not determined by consensus protocols alone.
Rather, they co-evolve with design choices about membership, communication, committee formation, block production, voting, and finality.
These properties shape the conditions under which construct--metric combinations are applicable and interpretable.

\paragraph{Permission models}
Blockchain systems can be \emph{permissionless}, where consensus participation is open, or \emph{permissioned}, where participation is restricted to authenticated members under predefined rules~\cite{wang2019survey, kannengiesser2020trade}.
Permission models affect \ac{DoD} because they define who can participate in consensus-critical roles and under which conditions.

\paragraph{Finality models}
Finality models specify when appended blocks or ordered transactions are considered finalized, that is, when they cannot be reverted under the assumptions of the protocol~\cite{kannengiesser2020trade}.
With \emph{deterministic finality}, a committed block or transaction is final once agreement is reached.
With \emph{probabilistic finality}, the probability of reversal decreases as additional blocks are appended.
Finality models affect scalability and security because stronger finality often requires additional communication, voting, or waiting time.

\paragraph{Fault models}
Blockchain systems differ in the faults and attacks they assume.
Some systems are designed for crash faults, whereas others address Byzantine or economically motivated adversaries~\cite{lamport1982byzantine, fischer1985flp, gervais2016security, eyal2018selfish}.
The relevant fault model is protocol- and layer-specific.
For example, an ordering service may provide crash-fault tolerance, while the surrounding blockchain system may rely on additional validation, replication, or economic assumptions to address Byzantine behavior.

\paragraph{Performance}
Consensus protocols influence transaction throughput and confirmation latency~\cite{kannengiesser2020trade, xiao2020survey, esmaili2025performance}.
Voting-based protocols with deterministic finality can incur increasing communication overhead as the number of validating nodes grows, whereas protocols with probabilistic finality can support open participation with weaker coordination requirements per decision but may require waiting for multiple confirmations.

\subsection{Blockchain Trilemma Dimensions and Their Interrelationships}
\label{sec:back-blockchain-trilemma}

We use the blockchain trilemma as an analytical lens for comparing blockchain system designs with respect to \ac{DoD}, scalability, and security.
The trilemma is related to established distributed-system theorems such as the CAP theorem, but it is not identical to them.
CAP formalizes a limitation for replicated distributed systems under network partitions: a system cannot simultaneously guarantee consistency and availability in the presence of partitions~\cite{brewer2012cap, gilbert2002brewer}.
Although this limitation also applies to blockchain systems as replicated distributed systems~\cite{wang2024gbt, li2025principles}, CAP does not cover several concerns that shape blockchain-system design, including distribution of decision power among validating nodes and adversarial control over consensus resources.
We therefore treat CAP as a related but narrower distributed systems theorem and use the blockchain trilemma as an analytical lens for blockchain system design tensions.

\begin{figure*}[!t]
    \centering
    \includegraphics[width=0.9\textwidth]{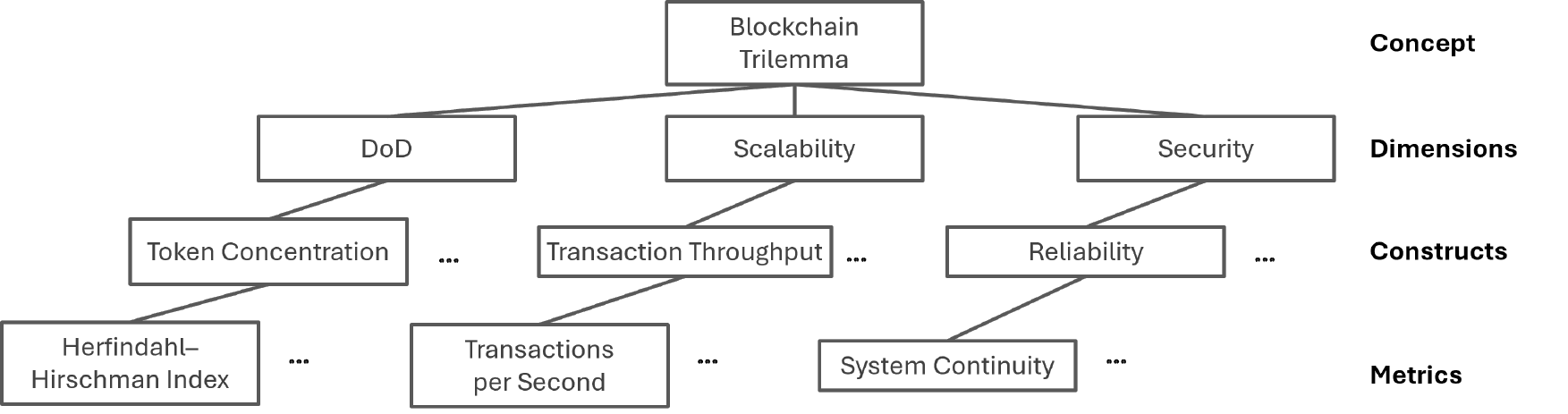}
    \caption{Interrelationships between the blockchain trilemma, dimensions, constructs, and metrics.}
    \label{fig:blockchain-trilemma-organogram}
\end{figure*}

For measurement, we treat the blockchain trilemma as the overarching concept that organizes dimensions, constructs, and metrics.
As shown in Figure~\ref{fig:blockchain-trilemma-organogram}, the trilemma is not measured directly but analyzed through the dimensions \ac{DoD}, scalability, and security.
Each dimension can be analyzed through constructs, each of which captures a specific measurable aspect of that dimension.
A \emph{construct} is a measurable aspect of a trilemma dimension in a blockchain system design.
Constructs are operationalized through metrics.
A \emph{metric} assigns values to objects or events based on specified \emph{input parameters} and provides an observable basis for analyzing a blockchain system with respect to a construct.

\subsubsection{Blockchain Trilemma Dimensions}

The blockchain trilemma highlights the recurring difficulty of simultaneously improving \ac{DoD}, scalability, and security in blockchain systems that are consistent with the system model described in Section~\ref{sec:foundations-of-bc-technology}.
In the following, we synthesize extant definitions of the three blockchain trilemma dimensions and clarify their scope.

\paragraph{Degree of Decentralization}
\Ac{DoD} has been approached from multiple perspectives: equitable and independent participation in consensus~\cite{kannengiesser2020trade, zhang2022sok, wuwang2026drivers}, geographical and network diversity~\cite{xiao2020survey, mssassi2024blockchain}, and resource distribution~\cite{juodis2024overview, nakai2024formulation, quattrocchi2024blockchain, jia2022measuring}.
We refer to \textit{\ac{DoD} as the extent to which validating nodes have equitable influence on consensus and exercise this influence (quasi-)autonomously}~\cite{wuwang2026drivers}.
Validating nodes exercise influence on consensus through consensus-critical tasks.
Geographic distribution and economic concentration are treated as influencing factors only if they shape equity of influence and autonomy of validating nodes.

\paragraph{Scalability}
Scalability is commonly associated with the maximum transaction processing rate, that is, transaction throughput~\cite{dinh2017blockbench, ahmad2021performance, mssassi2024blockchain, quattrocchi2024blockchain}.
Other works define scalability more broadly as the efficiency of a blockchain system as scale and workload increase~\cite{xu2023survey, gramoli2023diablo}.
Scalability can also depend on how transaction throughput and confirmation latency change with the number of validating nodes~\cite{graebe2020metrics, kannengiesser2020trade}.
We refer to \textit{scalability as the ability of a blockchain system to handle changing workloads and system sizes, including changes in transaction volume and validating-node count involved in consensus-critical tasks}~\cite{voron2023planetary, xiao2020survey, monte2020scaling}.

\paragraph{Security}
Security has likewise been defined inconsistently in blockchain-trilemma literature.
Some works emphasize availability and fault tolerance~\cite{dinh2017blockbench, kannengiesser2020trade, ren2023bbsf}, others relate security to partition tolerance and CAP-style limitations~\cite{brewer2012cap, gilbert2002brewer, wang2024gbt, li2025principles}, while others stress resistance against malicious attacks~\cite{eyal2018selfish, gervais2016security}.
We harmonize these perspectives by referring to \textit{security as the degree to which the blockchain system preserves correct operation and continued progress under faults, partitions, and malicious attacks}.
This scope follows the blockchain system model defined in Section~\ref{sec:foundations-of-bc-technology} and excludes application-layer artifacts such as smart-contract vulnerabilities.

This definition treats security as an umbrella dimension, as is common in blockchain-trilemma literature~\cite{kannengiesser2020trade, werth2023review}, while keeping distinct distributed-systems concepts separate.
In distributed-systems terminology, safety concerns the avoidance of conflicting, invalid, or inconsistent outcomes, whereas liveness concerns continued progress~\cite{alpern1985defining, alpern1987recognizing}.
These properties are therefore not treated as security constructs in our catalog.
Rather, they clarify how fault-tolerance and attack-related analyses can provide evidence for the security dimension: such analyses fall within our security scope when they examine whether faults, partitions, or malicious behavior compromise safety, liveness, or both~\cite{lamport1982byzantine, castro1999practical, eyal2018selfish, gervais2016security}.

\subsubsection{Interrelationships Between the Blockchain Trilemma Dimensions}
\label{sec:back-interrelationships}

The three blockchain trilemma dimensions are analytically separable but not independent.
The following paragraphs summarize the main pairwise tensions among \ac{DoD}, scalability, and security, while emphasizing that their direction and strength depend on the blockchain system design.

\paragraph{DoD vs.~Scalability}
A blockchain system can increase \ac{DoD} by reducing the concentration of influence on consensus among validating nodes.
If this requires more validating nodes to exercise influence in each consensus decision, communication and coordination overhead can increase.
This can increase confirmation latency or reduce transaction throughput and thereby decrease scalability~\cite{kannengiesser2020trade, gramoli2022blockchain, normann2025tradeoff}.
Conversely, scalability increases when blockchain systems reduce communication and coordination overhead.
This may be achieved by assigning consensus-critical tasks to fewer validating nodes per decision, for example through leaders or committees.
However, this decreases \ac{DoD} when the assignment repeatedly privileges a fixed or otherwise dominant subset of validating nodes, so that influence on consensus is no longer equitable among validating nodes.
Rotating leaders or committees can improve scalability without necessarily reducing \ac{DoD}, but only if they preserve equitable influence on consensus over time; the Ethereum system's slot-based proposers and attesting committees illustrate this design distinction~\cite{ethereum2025arch}.

\paragraph{DoD vs.~Security}
Effects of \ac{DoD} on security can be two-sided.
On the one hand, when influence on consensus is less concentrated, adversaries must compromise, corrupt, or control a larger share of consensus-relevant influence to affect correct operation or continued progress of the blockchain system.
This can increase the cost of attack and reduce the likelihood of collusion, resource dominance, or consensus capture~\cite{kannengiesser2020trade, gervais2016security}.

On the other hand, mechanisms that make influence on consensus less concentrated can decrease security when they weaken the propagation or coordination assumptions required by the consensus protocol.
For example, slower block propagation can increase forks, stale blocks, or temporary inconsistencies.
In blockchain systems with probabilistic finality, such effects can enlarge the window for double-spending or selfish-mining attacks~\cite{decker2013propagation, eyal2018selfish, gobel2017increased, gervais2016security}.
Network-level attacks, such as eclipse or routing attacks, fall within this scope when they affect transaction propagation, consensus participation, ordering, finality, or continued progress of the core blockchain system.

Conversely, security can increase when blockchain systems strengthen coordination among validating nodes or restrict participation to preserve correct operation and continued progress under the consensus protocol's fault and network assumptions.
Permissioned membership, leader-based ordering, or committees can support this by limiting which validating nodes participate in ordering, block proposal, attestation, voting, or block validation.
However, these mechanisms decrease \ac{DoD} when they make influence on consensus less equitable or limit the quasi-autonomy with which validating nodes exercise this influence~\cite{castro1999practical, hyperledger2025ordering, wuwang2026drivers}.

\paragraph{Scalability vs.~Security}
Scalability increases when blockchain systems process more transactions per time unit or confirm transactions faster.
This can be achieved by reducing communication among validating nodes, assigning consensus-critical tasks to fewer validating nodes per decision \cite{kannengiesser2020trade}.
However, these mechanisms can decrease security when correct operation or continued progress depends on fewer validating nodes or on assumptions that are harder to preserve under faults, partitions, or malicious attacks.
If these validating nodes crash, become overloaded, are partitioned, or are compromised, transaction propagation, ordering, validation, or voting can fail or slow down, interrupting correct operation or continued progress of the blockchain system.

Scalability can also increase when blockchain systems create or confirm blocks more quickly.
If block propagation does not keep pace with block creation, forks and stale blocks become more likely.
In blockchain systems with probabilistic finality, such temporary inconsistencies can enlarge the window for double-spending or selfish-mining attacks~\cite{decker2013propagation, eyal2018selfish, gervais2016security, kannengiesser2020trade}.

Security-oriented designs often add redundancy, communication, or waiting time. These mechanisms can improve fault and attack tolerance, but they also tend to reduce scalability by lowering throughput or increasing confirmation latency~\cite{kannengiesser2020trade, xiao2020survey, gramoli2023diablo}.
The Solana system illustrates one instantiation of this tension: leader scheduling and high-performance networking support high transaction throughput, while fault-injection studies indicate that continued progress can become sensitive to validating-node failures or overload~\cite{solana2019update, gramoli2025stabl}.

\subsection{Related Research on Measuring Blockchain Systems Through the Blockchain Trilemma}
\label{sec:approaches-to-bc-system-behavior}

Existing knowledge about measuring blockchain-system properties associated with the blockchain trilemma can be organized around three research perspectives: conceptual, formal, and empirical.
The conceptual perspective defines or qualitatively compares \ac{DoD}, scalability, and security across blockchain systems, consensus protocols, and architectural design choices~\cite{xiao2020survey, xu2023survey, kannengiesser2020trade, werth2023review, reno2025navigating}.
This perspective establishes the blockchain trilemma as a useful lens for blockchain-system design, but definitions of the three dimensions differ across studies.
For example, \ac{DoD} is sometimes associated with network size, sometimes with geographical or network diversity, and sometimes with autonomy, resource distribution, or equitable participation in consensus~\cite{xu2023survey, xiao2020survey, kannengiesser2020trade, zhang2022sok, werth2023review, ovezik2025sok}.
Such definitional heterogeneity limits comparability because studies may use the same blockchain trilemma dimension to refer to different measurable phenomena.

The formal perspective models trilemma-related properties or derives relationships under explicit assumptions~\cite{nakai2024formulation, mssassi2024blockchain}.
This perspective provides conceptual precision by specifying relationships among variables, resource distributions, or system assumptions.
For example, formal and analytical work has examined decentralization metrics, resource-distribution measures, and trilemma-related trade-offs.
However, such results are often tied to specific modeling assumptions, consensus mechanisms, or individual dimensions, which limits their direct use as general guidance for selecting constructs and metrics across heterogeneous blockchain systems.

The empirical perspective applies selected metrics to concrete blockchain systems, datasets, simulations, or benchmark environments~\cite{dinh2017blockbench, ren2023bbsf, gramoli2023diablo, fu2024quantifying, quattrocchi2024blockchain, ovezik2025sok}.
This perspective shows that blockchain-system properties associated with the trilemma dimensions can be quantified in concrete settings, for example, through token concentration, wealth distribution, confirmation latency, transaction throughput, fault tolerance, \ac{SBR}, or cost of attack.
However, empirical and benchmarking studies typically use metrics selected for their respective systems, datasets, or experimental setups.
As a result, they provide valuable evidence for particular operationalizations, but they do not by themselves provide a consolidated account of how constructs and metrics compare across the blockchain trilemma dimensions.

Across these perspectives, existing research provides definitions, formalizations, and selected measurements of blockchain-system properties associated with \ac{DoD}, scalability, and security.
However, defensible metric selection and comparable blockchain-system evaluation require a mapping from blockchain trilemma dimensions to constructs, metrics, and applicability conditions within a common system boundary.
Existing studies provide limited guidance on how constructs and metrics relate to the blockchain trilemma dimensions, under which blockchain system designs they apply, and how their values should be interpreted.
Consequently, researchers and practitioners can often determine how individual metrics are computed, but receive limited support in assessing when these metrics can be meaningfully applied and compared across blockchain systems.
Addressing this gap is the objective of the present study.
Table~\ref{tab:related-work-comparison} positions representative studies relative to this synthesis problem.

\begin{table*}[!t]
\centering
\caption{Positioning of representative studies on measuring blockchain-system properties associated with the blockchain trilemma dimensions.
\acs{DoD} = \acl{DoD}, Sc = scalability, Se = security, CM = construct--metric synthesis, IV = input parameters/data requirements, AC = applicability conditions, AA = analysis approaches, CS = cross-system comparability.}
\label{tab:related-work-comparison}
\renewcommand{\arraystretch}{0.92}
\setlength{\tabcolsep}{1.5pt}
\small

\begin{tabularx}{\textwidth}{
>{\raggedright\arraybackslash}m{2.8cm}
>{\centering\arraybackslash}m{0.55cm}
>{\centering\arraybackslash}m{0.55cm}
>{\centering\arraybackslash}m{0.55cm}
>{\centering\arraybackslash}m{0.65cm}
>{\centering\arraybackslash}m{0.65cm}
>{\centering\arraybackslash}m{0.65cm}
>{\centering\arraybackslash}m{0.65cm}
>{\centering\arraybackslash}m{0.65cm}
>{\raggedright\arraybackslash}X
}
\toprule
\textbf{Reference} 
& \textbf{DoD} 
& \textbf{Sc} 
& \textbf{Se} 
& \textbf{CM} 
& \textbf{IV} 
& \textbf{AC} 
& \textbf{AA} 
& \textbf{CS} 
& \textbf{Primary contribution} \\
\midrule

Kannengießer et al.~\cite{kannengiesser2020trade}
& \full & \full & \full & \none & \none & \partly & \none & \partly
& Conceptualizes trade-offs among blockchain dimensions. \\

Xiao et al.~\cite{xiao2020survey}
& \partly & \full & \full & \none & \none & \partly & \none & \partly
& Surveys consensus protocols and related dimensions. \\

Xu et al.~\cite{xu2023survey}
& \partly & \full & \full & \none & \none & \partly & \none & \partly
& Reviews blockchain systems and consensus-related dimensions. \\

Werth et al.~\cite{werth2023review}
& \full & \full & \full & \partly & \none & \partly & \none & \partly
& Qualitatively compares blockchain systems with respect to the blockchain trilemma. \\

Reno and Roy~\cite{reno2025navigating}
& \partly & \full & \full & \partly & \none & \partly & \none & \partly
& Surveys architectures and trilemma-related trade-offs. \\

Zhang et al.~\cite{zhang2022sok}
& \full & \none & \partly & \partly & \partly & \partly & \none & \partly
& Formalizes and compares \ac{DoD} metrics. \\

Nakai et al.~\cite{nakai2024formulation}
& \full & \full & \partly & \partly & \partly & \full & \partly & \partly
& Formalizes selected trilemma relationships for \ac{PoW} blockchains under modeling assumptions. \\

Mssassi et al.~\cite{mssassi2024blockchain}
& \full & \full & \full & \partly & \partly & \partly & \none & \partly
& Discusses formal relationships among the blockchain trilemma dimensions. \\

Dinh et al.~\cite{dinh2017blockbench}
& \none & \full & \partly & \partly & \full & \partly & \full & \partly
& Provides benchmark metrics for measuring blockchain system performance. \\

Ren et al.~\cite{ren2023bbsf}
& \none & \full & \full & \partly & \full & \partly & \full & \partly
& Benchmarks based on performance and fault-tolerance-related metrics. \\

Gramoli et al.~\cite{gramoli2023diablo}
& \none & \full & \partly & \partly & \full & \partly & \full & \partly
& Provides benchmarking infrastructure for scalability analysis. \\

Fu et al.~\cite{fu2024quantifying}
& \full & \full & \partly & \partly & \partly & \partly & \partly & \partly
& Quantifies selected blockchain system properties for \ac{PoS} systems. \\

Quattrocchi et al.~\cite{quattrocchi2024blockchain}
& \full & \full & \full & \partly & \partly & \partly & \partly & \partly
& Applies selected metrics to blockchain system properties associated with the trilemma dimensions. \\

Ovezik et al.~\cite{ovezik2025sok}
& \full & \none & \none & \partly & \partly & \partly & \partly & \partly
& Systematizes decentralization measurement workflows and evaluates metric choices. \\

\textit{This study}
& \full & \full & \full & \full & \full & \full & \full & \full
& Synthesizes constructs, metrics, input parameters, applicability conditions, limitations, and analysis approaches. \\

\bottomrule
\end{tabularx}

\vspace{0.3em}
\begin{minipage}{\textwidth}
\footnotesize
\full~= addressed in detail; \partly~= partially addressed; \none~= not addressed or not central.
The coding positions each contribution relative to the scope of this study; it does not imply that prior work is incomplete for its own purpose.
\end{minipage}
\end{table*}
    \section{Research Approach}
\label{sec:research-approach}

We identified constructs, metrics, input parameters, and applicability conditions for analyzing blockchain system designs through the lens of the blockchain trilemma in four steps. First, we constructed the literature corpus through a concept-centric systematic literature search~\cite{webster2002analyzing} and supplementary purposive sampling~\cite{palinkas2015purposeful}. Second, we used the blockchain trilemma dimensions as a deductive coding frame and inductively synthesized constructs and metrics from the reviewed literature~\cite{braun2006using,thomas2008methods}. Third, we reanalyzed the corpus to derive applicability conditions for each construct--metric combination. Fourth, we sought external author feedback on selected construct descriptions, metric specifications, and mappings.

\subsection{Literature Search}
\label{sec:literature-search}

We conducted a concept-centric systematic literature search~\cite{webster2002analyzing} to identify publications that analyze blockchain systems with respect to the blockchain trilemma, its dimensions, or related benchmarking and measurement concerns. To evaluate the relevance of publications, we applied five inclusion criteria: \textit{English language}, \textit{level of detail}, \textit{peer-reviewed}, \textit{topic fit}, and \textit{uniqueness} (see Table~\ref{tab:inclusion-criteria}).

\begin{table*}[!t]
\renewcommand{\arraystretch}{1.3}
\caption{Inclusion criteria used in the literature search.}
\label{tab:inclusion-criteria}
\centering
\footnotesize
\begin{tblr}{
width = \textwidth,
vlines = {},
hlines = {},
column{1} = {2.6cm},
column{2} = {co=1}
}
\textbf{Criterion}  & \textbf{Description}\\
English Language
& The publication must be in English. \\
Level of Detail
& The publication must present sufficient descriptions and explanations of the investigated blockchain trilemma dimension(s), the construct(s) used to analyze them, and the metric(s) used to operationalize these constructs. \\
Peer-Reviewed
& The publication must be peer-reviewed. \\
Topic Fit
& The publication must focus on analyzing at least one blockchain trilemma dimension through constructs and associated metrics. The construct(s) or metric(s) must apply to our system model and must not depend on specialized hardware, such as trusted execution environments, application-layer artifacts, such as smart contracts, or peripheral software artifacts, such as state channel networks. \\
Uniqueness
& The publication must be the latest available version and must not be a duplicate in the literature set. \\
\end{tblr}

\end{table*}

\paragraph{Initial systematic search.}
The initial search established the core literature corpus. We searched ACM Digital Library, IEEE Xplore, ScienceDirect, and Scopus on March 26, 2024, using the search string \textit{(`benchmarking' AND `blockchain trilemma')}. The search was restricted to publications from 2014 to 2024. These databases were selected because they provide broad coverage of computer science, information systems, and engineering research relevant to blockchain-system evaluation. The search string was designed as a high-precision anchor search to identify publications that explicitly connect benchmarking or measurement with the blockchain trilemma.

The search returned 1,814 potentially relevant publications: 1,258 from ACM Digital Library, 546 from IEEE Xplore, 7 from ScienceDirect, and 3 from Scopus. We screened all 1,814 publications based on title, keywords, and abstract against the inclusion criteria. This step excluded 436 publications: 4 were not in English, 348 were not peer-reviewed, 77 lacked topic fit, and 7 were duplicates. This left 1,378 records for full-text assessment. Screening decisions were discussed among the authors until agreement was reached. Because the objective of the screening procedure was consensual relevance assessment rather than the aggregation of independent ratings, we used a consensus-based procedure and did not calculate inter-rater reliability statistics such as Cohen's kappa.

We subsequently performed a full-text assessment of the remaining 1,378 publications using the same inclusion criteria. During this stage, 1,210 publications were excluded because they did not provide sufficient detail, and 24 additional publications were excluded because they lacked topic fit. The full-text assessment yielded \textit{144 relevant publications}.

\paragraph{Supplementary purposive search.}
During the thematic synthesis (Section~\ref{sec:construct-metric-synthesis}), we observed that constructs related to \ac{DoD} and security were underrepresented in the initial corpus. To improve conceptual coverage, we complemented the systematic database search with purposive sampling of information-rich publications~\cite{palinkas2015purposeful}. The supplementary search targeted publications that addressed conceptual blind spots, operationalized underrepresented blockchain trilemma dimensions, introduced relevant metrics, analyzed blockchain system designs not sufficiently covered in the initial corpus, or reflected more recent blockchain system designs, such as Ethereum~2.0 rather than Ethereum~1.0.

We conducted the supplementary search using Google Scholar by re-applying the original search string and following citation trails from information-rich publications. This step was purposive rather than exhaustive. Its objective was to identify relevant studies that may have been omitted from the initial database search because of indexing limitations, terminology differences, or the evolving nature of blockchain-system research. Retrieved records were screened for conceptual relevance, uniqueness, and fit with the same inclusion and exclusion criteria defined in Table~\ref{tab:inclusion-criteria}. Duplicates and studies previously assessed during the initial screening phase were excluded. We did not record a complete count of all Google Scholar results screened; accordingly, we treat this step as a coverage-oriented supplement rather than as a fully reproducible exhaustive search. This supplementary purposive search resulted in the inclusion of \textit{17 additional publications}, yielding an intermediate corpus of \textit{161 studies}.

\paragraph{Additional database search.}
To update and broaden the corpus during revision, we conducted an additional database search in Web of Science and DBLP on April 26, 2026. These databases provide complementary coverage to the previously searched sources. The search returned 128 publications: 106 from Web of Science and 22 from DBLP. These publications were screened according to the inclusion criteria in Table~\ref{tab:inclusion-criteria}. Of the 128 publications, 39 were excluded due to insufficient detail, 70 lacked topic fit, 1 was not in English, and 3 were duplicates. In addition, 11 publications had already been retrieved from previously searched databases. The additional database search therefore added \textit{4 relevant publications}. The final corpus comprised \textit{165 studies}.

\subsection{Construct and Metric Synthesis}
\label{sec:construct-metric-synthesis}

We analyzed the final corpus using framework-guided deductive--inductive thematic synthesis~\cite{braun2006using,thomas2008methods}. The blockchain trilemma and its three dimensions---\ac{DoD}, scalability, and security---provided the deductive top-level coding frame. Within these categories, we inductively coded constructs and metrics used in the reviewed literature. A construct was coded when a publication used a recurring measurable aspect to analyze at least one blockchain trilemma dimension. A metric was coded when a publication specified how such an aspect was operationalized, calculated, or applied.

The coding scheme was refined iteratively by moving between the literature corpus, the definitions introduced in Section~\ref{sec:back-blockchain-trilemma}, and the emerging construct and metric set. Initial codes, such as \textit{block-proposal randomness}, \textit{fault tolerance}, and \textit{throughput}, were derived from the terminology and measurement approaches used in the reviewed publications. We then compared, merged, renamed, or separated codes where publications used different terms for conceptually similar phenomena or similar terms for different phenomena.

We defined a \textit{construct} as a recurring concept used in the literature to operationalize at least one blockchain trilemma dimension and associated with at least one metric. To reduce redundancy, overlapping constructs were merged. For example, \textit{robustness} was subsumed under \textit{fault tolerance}, and \textit{availability}- and \textit{success-rate}-related codes were grouped under \textit{transaction confirmation ratio} as a scalability construct. Similarly, repeatedly occurring maximum possible throughput- and confirmation-latency-related codes were consolidated into broader constructs. This consolidation explains the reduction from the initial code set to the final construct set.

To enhance transparency and trustworthiness of the thematic synthesis~\cite{nowell2017thematic}, subsets of the literature were coded independently and discrepancies were resolved through discussion. The coding scheme was refined until agreement was reached on the interpretation, naming, and assignment of constructs and metrics. This process resulted in a stable structure of 12 constructs distilled from 417 initial codes.

Each construct was mapped to one of the blockchain trilemma dimensions based on patterns in the literature and alignment with the definitions introduced in Section~\ref{sec:back-blockchain-trilemma}. The mapping, therefore, did not rely only on how a publication labelled a construct. Instead, we assessed whether the construct captured equitable influence on consensus and quasi-autonomy, workload- and size-dependent performance, or preservation of correct operation and continued progress under faults, partitions, and malicious attacks. For instance, \textit{transaction throughput} and \textit{confirmation latency} were assigned to scalability, while \textit{fault tolerance} and \ac{SBR} were associated with security. Mapping disagreements were resolved through collaborative review and consensus-based discussion.

Metric specifications were then compared across publications to identify equivalent equations, alternative operationalizations, required input parameters, and differences in interpretation. Candidate constructs or metric specifications were excluded when their conceptual mapping, equation, or reported calculation could not be reconciled with the stated construct and the issue could not be resolved from the source text. Where necessary, we contacted the original authors for clarification. If the issue remained unresolved, we excluded the candidate construct or metric specification from the final results to preserve conceptual clarity and empirical traceability.

In the final coding iterations, additional coding within the selected corpus no longer yielded substantially new constructs or associated metrics. We therefore considered the coding structure sufficiently stable for synthesizing how prior literature analyzes the blockchain trilemma dimensions through constructs and operationalizes these constructs through metrics.

\subsection{Applicability Condition Identification}
\label{sec:applicability-condition-analysis}

We analyzed the literature corpus to identify applicability conditions for each construct-metric combination. Applicability conditions describe when a construct and metric can be meaningfully used in analyses.
We distinguished conditions related to \emph{blockchain system design} (when does the metric apply to the design?) from conditions related to the \emph{analysis approach} (what information is needed to use the metric?).

For applicability conditions arising from blockchain system design, we extracted which blockchain systems had been analyzed with which metrics and compared the system properties that affected metric applicability. We examined, for example, whether the measured resource or task-shaped influence on consensus, whether tokens were consensus-relevant, whether validating-node roles could be identified, whether the system used deterministic or probabilistic finality, and whether the metric presupposed specific block creation or propagation behavior. We then synthesized recurring similarities and differences across blockchain system designs into applicability conditions.

For applicability conditions arising from analysis approaches, we examined how publications obtained the input parameters required by each metric. We coded whether data were collected from empirical blockchain observations, configured in simulations, derived analytically, or measured in benchmarking environments. We then assessed which data collection and evaluation conditions had to be satisfied for the metric to be methodologically meaningful, such as observing the relevant time window, applying a specified workload, using a defined finality criterion, or attributing resources to the entities that exercise validating-node influence. The resulting applicability conditions therefore connect each construct--metric combination to both the blockchain system design being analyzed and the analysis approach used to obtain the required data.

\subsection{External Feedback on Construct Descriptions, Metrics, and Mappings}
\label{sec:external-feedback}

After completing the literature analysis, we sought external feedback on construct descriptions, metric specifications, and mappings to the blockchain trilemma dimensions. We contacted author teams whose publications were included in the corpus and were central to the extracted constructs or metrics. In total, we reached out to 29 author teams and asked them to review the parts of the synthesis related to their work. We sent the relevant construct descriptions, metric specifications, and mappings in a PDF version of the manuscript, invited written feedback via email, and offered virtual meetings for discussion.

Thirteen author teams responded. Their feedback led to refinements of construct descriptions, metric specifications, and mappings. For example, we refined \textit{confirmation latency} to better reflect the design of Ethereum~2.0, which was not fully represented in the analyzed literature, and refined the throughput construct to distinguish empirical transaction throughput from capacity-bound estimates. To remain consistent with the scope of this work, we also rephrased \textit{trusted third party} as \textit{validating node}. Two author teams identified minor errors in metric equations, which we corrected. Most responding author teams agreed with the descriptions and mappings or provided suggestions for future extensions. The external feedback step therefore supported the refinement of the construct descriptions, metric specifications, and mappings.
    \section{Constructs, Metrics, and Analysis Approaches}
\label{sec:results}

This section presents the constructs and metrics identified for measuring blockchain-system properties associated with the blockchain trilemma dimensions. The blockchain trilemma provides the analytical lens, whereas the object of analysis is the blockchain system design within the core blockchain-system boundary defined in Section~\ref{sec:foundations-of-bc-technology}. Subsection~\ref{sec:results-constructs} synthesizes the constructs, metrics, and caveats; Subsection~\ref{sec:results-applicability-conditions} summarizes applicability conditions and analysis approaches.

\subsection{Constructs and Metrics for Measuring Blockchain-System Properties Associated with the Blockchain Trilemma Dimensions}
\label{sec:results-constructs}

We identified 12 constructs associated with the blockchain trilemma dimensions (see Table~\ref{tab:output-variables}): four for \ac{DoD}, three for scalability, and five for security.
These constructs are operationalized through 15 metrics; normalizations and auxiliary equations support interpretation but are not counted as separate metrics.
The input parameters to the metrics are summarized in Table~\ref{tab:input-parameters} in Appendix~\ref{appdx:ovw-input-parameters}.

\begin{table*}[!t]
\renewcommand{\arraystretch}{1.3}
\caption{Overview of the identified constructs associated with the blockchain trilemma dimensions: \ac{DoD}, scalability, and security.}
\label{tab:output-variables}
\centering
\footnotesize
\begin{tblr}{
    width = \textwidth,
    rows = {valign = m},
    cell{2}{1} = {r=4}{valign = m, halign = c},
    cell{6}{1} = {r=3}{valign = m, halign = c},
    cell{9}{1} = {r=5}{valign = m, halign = c},
    column{1} = {0.8cm},
    column{2} = {4cm},
    column{3} = {co=1},
    hline{1} = {2-3}{},
    hline{2-Z} = {1-Z}{},
    vline{1} = {2-Z}{},
    vline{2-Z} = {1-Z}{}
}

    & \textbf{Construct}
    & \textbf{Description} \\

    \rotatebox[origin=c]{90}{\textbf{Degree of Decentralization}}
    & Block-Proposal Randomness
    & The uncertainty with which block-proposal influence is distributed across validating nodes. \\

    & Consensus Resource Distribution
    & The extent to which consensus-relevant resources are distributed across validating nodes or entities that control validating-node influence on consensus. \\

    & Token Concentration
    & The concentration of consensus-relevant token shares attributed to validating nodes. \\

    & Wealth Distribution
    & The inequality of consensus-relevant token amounts attributed to validating nodes. \\

    \rotatebox[origin=c]{90}{\textbf{Scalability}}
    & Confirmation Latency
    & The time elapsed until a submitted transaction or proposed block is confirmed or finalized under the reported finality criterion. \\

    & Transaction Confirmation Ratio
    & The share of valid submitted transactions confirmed within the observation window under a specified workload and finality criterion. \\

    & Transaction Throughput
    & The number of confirmed transactions processed per time unit under a specified workload or capacity-bound assumption. \\

    \rotatebox[origin=c]{90}{\textbf{Security}}
    & Cost of Attack
    & The estimated fiat-denominated cost required to obtain sufficient consensus-relevant resources to execute a specified attack on a blockchain system. \\

    & Fault Tolerance
    & The extent to which a blockchain system preserves correct operation and continued progress under a specified fault model. \\

    & Reliability
    & The probability that the blockchain system continues to provide correct service without failure over a specified time interval. \\

    & Stale Block Rate
    & The ratio of valid propagated blocks that are not included in the canonical chain to confirmed canonical-chain blocks during an observation window. \\

    & Success Probability of Selfish Mining
    & The probability that a selfish validating node's private branch overtakes or manipulates the honest public branch and becomes accepted by the system. \\

\end{tblr}
\end{table*}


\subsubsection{Degree of Decentralization} 
\label{sec:results-dod-constructs}

To estimate \ac{DoD}, we identified four constructs that capture distinct aspects of how influence on consensus is allocated and exercised in blockchain system designs: \textit{block-proposal randomness}, \textit{consensus resource distribution}, \textit{token concentration}, and \textit{wealth distribution}.
These constructs approximate aspects of \ac{DoD} only when the measured distribution affects equitable influence on consensus or the autonomous exercise of that influence by validating nodes.
Where ownership or operational control differs from validating-node identity, node-level measurements should be interpreted together with the reported attribution assumptions.

\constructDefinition{Block-Proposal Randomness}{The uncertainty with which block-proposal influence is distributed across validating nodes.}

Depending on the consensus protocol, block proposal may be assigned to an elected leader, as in Raft~\cite{ongaro2014raft}, or to a (pseudo-)randomly selected validating node, as in Nakamoto-style or Gasper-style consensus~\cite{nakamoto2008bitcoin, neu2022failure}. Block-proposal randomness is commonly computed using Shannon entropy~\cite{wu2019information, lin2021measuring, jia2022measuring, juodis2024overview, liu2023incentive, ovezik2025sok, yan2025data, chemaya2025dataset, gochhayat2020measuring}, which measures uncertainty in discrete outcomes~\cite{shannon1948mathematical}, here the observed distribution of block proposals across validating nodes:

\begin{equation}
\label{eq:BPR}
H \;=\; - \sum_{i=1}^{n} p_i \log_2 p_i,\qquad
p_i \;=\; \frac{b_i}{\sum_{j=1}^{n} b_j},
\end{equation}

where $n$ is the validating-node count and $b_i$ is the number of blocks proposed by validating node $i$ (with $0\log 0 := 0$).
Let $k=\big|\{i:\,p_i>0\}\big|$ denote the number of validating nodes that proposed at least one block in the observation window. Then $H\le \log_2 k$, achieving $\log_2 n$ only if all validating nodes have equal nonzero proposal probability.

For robustness or sensitivity analysis, block-proposal randomness may also be computed using Rényi entropy, a parameterized generalization of Shannon entropy~\cite{ju2025crypto}. This is an optional entropy variant and is not counted as a separate metric.

For $\alpha\ge 0$, $\alpha\neq 1$, Rényi entropy is defined as
\begin{equation}
\label{eq:BPR_renyi}
H_\alpha \;=\; \frac{1}{1-\alpha}\log_2\!\left(\sum_{i=1}^{n} p_i^{\alpha}\right),
\end{equation}
which converges to $H$ as $\alpha\to 1$; larger $\alpha$ emphasize proposer dominance, while smaller $\alpha$ emphasize participation by low-frequency proposers.
The parameter $\alpha$ is chosen by the practitioner and is not inferred from the data; it determines the sensitivity of the entropy measure to proposer concentration.

For cross-configuration comparability, a unitless normalization is useful:
\begin{equation}
\label{eq:BPR_norm}
H_{\mathrm{norm}} \;=\;
\begin{cases}
\displaystyle \frac{H}{\log_2 k}, & k\ge 2,\\[6pt]
0, & k<2,
\end{cases}
\end{equation}
which lies in $[0,1]$ and equals $1$ under perfectly even proposer probability across the $k$ active proposers.

For illustration, suppose three validating nodes $n_1$, $n_2$, and $n_3$ proposed $1$, $1$, and $8$ blocks, respectively, out of 10 total. The resulting probabilities are $p_1=\tfrac{1}{10}$, $p_2=\tfrac{1}{10}$, and $p_3=\tfrac{8}{10}$, yielding $H=0.922$\,bits from~\eqref{eq:BPR}. The maximum entropy for three equally likely proposers is $\log_2(3)=1.585$\,bits, so $H_{\mathrm{norm}}\approx0.58$. This indicates intermediate dispersion of block-proposal influence among active proposers; by itself, it does not determine overall \ac{DoD}.

\paragraph{Metric Caveats}
Entropy-based metrics capture only the dispersion of observed block-proposal outcomes. They do not show whether proposed blocks are finalized, whether validating nodes have comparable network positions, or whether other consensus-critical tasks are distributed similarly. Bandwidth, latency, and peering advantages can affect which proposed blocks are accepted~\cite{gramoli2022blockchain, xiao2020survey}. Analysts should therefore report the finality criterion and role structure and interpret proposer entropy as a task-specific indicator of proposal influence.

\constructDefinition{Consensus Resource Distribution}{The extent to which consensus-relevant resources are distributed across validating nodes or entities that control validating-node influence on consensus.}

Consensus in blockchain systems depends on protocol-defined resources, such as hashing power, stake, or voting weight.
When a small number of validating nodes or controlling entities hold a large share of the resource that determines consensus influence, the corresponding aspect of \ac{DoD} decreases.

The distribution of consensus-relevant resources can be estimated using the \textit{Nakamoto coefficient}, defined as the minimum entity count that together controls enough of a consensus-relevant resource to exceed a compromise threshold~\cite{nakai2024formulation, fu2024quantifying, quattrocchi2024blockchain, juodis2024overview, lin2021measuring, ovezik2025sok, yan2025data, chemaya2025dataset}:

\begin{equation}
\label{eq:CRD}
NC \;=\; \min \left\{k \in \{1, \dots, m\} : \sum_{i=1}^{k} p_i \geq \theta \right\}.
\end{equation}

Here, $p_i$ denotes the share of a consensus-relevant resource controlled by attribution unit $i$, ordered such that $p_1 \geq p_2 \geq \dots \geq p_m$, $m$ denotes the number of observed attribution units, and $\theta$ denotes the compromise threshold.

The attribution unit may be a validating node, mining pool, staking provider, validator operator, organization, or beneficial owner, depending on the analysis objective and data availability. 
The threshold $\theta$ depends on the blockchain system and the attack being analyzed. For example, \ac{PoW} systems such as Bitcoin often use a threshold of more than $50\%$ of hashing power. In \ac{PoS} and BFT-style systems, the relevant threshold is usually expressed in stake or voting weight and depends on the protocol and attack model.

A higher Nakamoto coefficient indicates that more attribution units would need to coordinate or collude to exceed the selected compromise threshold.
A lower Nakamoto coefficient indicates that the selected consensus-relevant resource is concentrated among fewer attribution units, increasing the risk of capture or coordinated manipulation~\cite{nakai2024formulation, fu2024quantifying}.
The implication for \ac{DoD} depends on whether the selected resource and attribution level accurately represent influence on consensus.

\paragraph{Metric Caveats}
The Nakamoto coefficient captures only the minimum number of attribution units required to exceed a selected threshold; it does not describe the full resource distribution. Results depend on the threshold, observation window, resource definition, and attribution level, and can differ across validating nodes, mining pools, staking providers, operators, or beneficial owners. Shared ownership or coordination can cause $NC$ to overestimate effective \ac{DoD}.
The metric should therefore be interpreted as concentration in a consensus-relevant resource.

\constructDefinition{Token Concentration}{The concentration of consensus-relevant token shares attributed to validating nodes.}

In many public blockchain systems, protocol-native tokens influence participation in consensus-critical tasks. We refer to the attributed amount of such tokens as a \textit{consensus-relevant token amount}. When these amounts are unevenly distributed, a small number of validating nodes or controlling entities may obtain disproportionate influence on consensus.

Token concentration is typically measured using the \ac{HHI}~\cite{juodis2024overview, nakai2024formulation, fu2024quantifying, ovezik2025sok, yan2025data, chemaya2025dataset}, a standard economic indicator of concentration~\cite{herfindahh2024justice, rhoades1993herfindahl}:

\begin{equation}
\label{eq:TC}
HHI_{prop} \;=\; \sum_{i=1}^{n} \left( \frac{\tau_i}{\tau_{total}} \right)^2,
\end{equation}

where $\tau_i$ denotes the consensus-relevant token amount attributed to validating node $i$, $\tau_{total}=\sum_{i=1}^{n}\tau_i$, and $n$ denotes the validating-node count.
$HHI_{prop}\in\left[\tfrac{1}{n},\,1\right]$ uses \emph{proportional} shares without 10{,}000-scaling. Competition practice often reports $HHI_{10k}=10{,}000\times HHI_{prop}$.

For comparability across systems with different $n$, a common normalization is
\begin{equation}
\label{eq:TC_NORM}
HHI_{\mathrm{norm}} \;=\; \frac{HHI_{prop} - \tfrac{1}{n}}{1 - \tfrac{1}{n}}
\;=\; \frac{n}{n-1}\,HHI_{prop} \;-\; \frac{1}{n-1},
\end{equation}
which maps equal shares to $0$ and monopoly to $1$. If using the 10{,}000-scaled variant, apply $HHI_{prop}=HHI_{10k}/10{,}000$ in~\eqref{eq:TC_NORM}.

A high $HHI$ indicates concentration in the selected consensus-relevant token amount; a low $HHI$ indicates more even dispersion. The implication for \ac{DoD} depends on whether the selected token amount actually affects influence on consensus. For example, with eleven analysis units each holding five tokens (total $55$), $HHI_{prop}=1/11\approx0.0909$, so $HHI_{10k}\approx909$ and $HHI_{\mathrm{norm}}=0$. By contrast, if one analysis unit holds $37$ tokens and ten hold $1.8$ each (still $55$ total), then $HHI_{prop}\approx0.4633$, $HHI_{10k}\approx4633$, and $HHI_{\mathrm{norm}}\approx(0.4633-0.0909)/(1-0.0909)\approx0.410$, indicating stronger concentration in the selected token amount.

\paragraph{Metric Caveats}
For \ac{HHI}, it is assumed that the relevant token amounts can be attributed to validating nodes or controlling entities. This may not hold if rewards are redistributed off-chain, if one entity controls multiple validating nodes, or if apparent validating nodes share an operator. The metric should therefore be interpreted as concentration in the selected consensus-relevant token amount.

\constructDefinition{Wealth Distribution}{The inequality of consensus-relevant token amounts attributed to validating nodes.}

In many public blockchain systems, validating nodes receive protocol-native token rewards for successful consensus-critical tasks. Over time, such rewards and other token allocations may accumulate unevenly. Wealth distribution captures how equally or unequally consensus-relevant token amounts are distributed across validating nodes or controlling entities. It provides insight into \ac{DoD} when the measured token amounts are relevant to consensus participation, influence on consensus, or reward accumulation in the analyzed blockchain system.

Wealth distribution is commonly measured using the \textit{Gini coefficient}~\cite{juodis2024overview, lin2021measuring, nakai2024formulation, fu2024quantifying, quattrocchi2024blockchain, jia2022measuring, ovezik2025sok, yan2025data, chemaya2025dataset, gochhayat2020measuring}, a standard inequality metric~\cite{gini1921measurement, dorfman1979formula, dagum1997new}.
It is defined as

\begin{equation}
\label{eq:WD_Gini}
Gini \;=\; \frac{\sum_{i=1}^{n} \sum_{j=1}^{n} \left| \tau_i - \tau_j \right|}{2 n \sum_{i=1}^{n} \tau_i},
\end{equation}

where $\tau_i$ and $\tau_j$ denote the consensus-relevant token amounts attributed to validating nodes $i$ and $j$, respectively, and $n$ denotes the validating-node count.

The Gini coefficient ranges from $0$ to $\frac{n-1}{n}$ for a finite validating node count, where $0$ denotes perfect equality and higher values denote stronger inequality. A higher value indicates stronger inequality in the consensus-relevant token amounts attributed to validating nodes.

The normalized $Gini$ is

\begin{equation}
\label{eq:WD_Gini_norm}
\mathrm{Gini_{\mathrm{norm}}} \;=\; Gini \cdot \frac{n}{n - 1},
\end{equation}

which maps the finite-sample range to $[0,1]$, where $0$ denotes equal token amounts across validating nodes and values close to $1$ denote high inequality.

Suppose five validating node $n_1, \ldots, n_5$ hold token amounts as follows: $n_1=2$, $n_2=3$, $n_3=5$, $n_4=6$, $n_5=2$. Substituting into equation~\ref{eq:WD_Gini} yields $Gini \approx 0.244$ and $\mathrm{Gini_{\mathrm{norm}}}\approx0.306$, indicating relatively low inequality in the selected token amount. By contrast, if one validating node holds all 18 tokens and the others none, $Gini = 0.8$, with $\mathrm{Gini_{\mathrm{norm}}} \approx 0.8 \cdot \frac{5}{5 - 1} \approx 1$, indicating maximal inequality for five validating nodes.

\paragraph{Metric Caveats}
The Gini coefficient assumes that consensus-relevant token amounts can be attributed to validating nodes. Node-level measurements may overestimate \ac{DoD} when one entity controls multiple validating nodes, when apparent validating nodes share an operator, or when rewards are redistributed off-chain.
It should therefore be interpreted as node-level inequality in the selected consensus-relevant token amount, not as a complete measure of \ac{DoD}.

\subsubsection{Scalability}
\label{sec:results-sca-constructs}

To estimate scalability, we identified three constructs: \textit{confirmation latency}, \textit{transaction confirmation ratio}, and \textit{transaction throughput}. These constructs should be interpreted together because each captures a different aspect of scalability.

\constructDefinition{Confirmation Latency}{The time elapsed until a submitted transaction or proposed block is confirmed or finalized under the reported finality criterion.}

Confirmation latency depends on the measurement boundary. In transaction-level benchmarks, it measures the time from transaction submission to confirmation or finalization. In block-level analyses, it measures the time from block proposal to confirmation or finalization. In blockchain systems with probabilistic finality, confirmation usually requires a specified number of subsequent blocks. In systems with deterministic finality, such as those using \ac{PBFT}, a block or transaction is finalized once the protocol's quorum and message-phase requirements are satisfied~\cite{castro1999practical}. The reported value therefore depends on the finality model, confirmation criterion, workload, and timestamp source.

Equation~\ref{eq:emp} gives a simple empirical estimate of confirmation latency $\mathrm{CL_{emp}}$ when start and confirmation or finalization timestamps are observable~(e.g.,~\cite{geyer2023end, nasrulin2022gromit, ren2023bbsf, wang2024gbt, alom2023blockmeter}):
\begin{equation}
\label{eq:emp}
\mathrm{CL_{emp}} \;=\; ConfirmTime - StartTime.
\end{equation}

Here, $StartTime$ denotes the transaction-submission timestamp in transaction-level benchmarks or the block-proposal timestamp in block-level analyses. $ConfirmTime$ denotes the timestamp at which the transaction or block satisfies the reported confirmation or finalization criterion.

In blockchain systems with probabilistic finality, confirmation latency can also be approximated from the required confirmation depth $b^\prime$ and the average block creation interval $BCI$~\cite{grabe2020not}:

\begin{equation}
\label{eq:conf}
\mathrm{CL_{conf}} \;=\; b^\prime \times BCI.
\end{equation}

$b^\prime$ denotes the required confirmation depth and $BCI$ denotes the average interval between consecutive canonical-chain blocks.

Equation~\ref{eq:epoch} models confirmation latency in the Ethereum system, where finality requires that a block be justified and then finalized in the next epoch~\cite{asgaonkar2024confirmation}:
\begin{equation}
\label{eq:epoch}
\mathrm{CL_{epoch}} \;=\; e_f \times TTJ,
\end{equation}
with
\begin{equation}
\label{eq:TTJ}
TTJ \;=\; e_l \times slot_{\mathit{time}},
\end{equation}
where $e_f$ is the number of epochs until finalization, $e_l$ is the number of slots per epoch, and $slot_{\mathit{time}}$ is the duration of a slot.

High confirmation latency typically indicates slower settlement and can reduce practical scalability, while low confirmation latency reflects faster confirmation or finalization. Equation~\ref{eq:emp} is broadly applicable when timestamps are observable, whereas equation~\ref{eq:conf} is specific to systems in which confirmation depends on a required number of subsequent blocks.

\paragraph{Metric Caveats}
All three metrics capture the time to confirmation, but ignore how many transactions are included per block. Block size, therefore, interacts with confirmation latency: larger blocks may contain more transactions but also prolong the time until confirmation \cite{ahmadjee2022study, gobel2017increased}. Hence, confirmation latency alone cannot fully represent scalability.

\constructDefinition{Transaction Confirmation Ratio}{The share of valid submitted transactions confirmed within the observation window under a specified workload and finality criterion.}

We use transaction confirmation ratio for the scalability-oriented metric that some benchmarking studies describe as availability or success rate~\cite{kannengiesser2020trade, gramoli2023diablo}. In this interpretation, the metric captures whether the core blockchain system confirms submitted valid transactions under load. It does not represent CAP availability in its full distributed-systems sense, because it does not by itself establish that every request to a non-failing node receives a response under arbitrary partitions. If the same ratio is measured during crash faults, Byzantine behavior, attacks, or network partitions, its value should be interpreted as security-related evidence about continued operation rather than as scalability under normal operation.

Transaction confirmation ratio is measured as the ratio of confirmed transactions $NumOfConfTr$ to submitted valid transactions $NumOfTr$ over a given observation period:

\begin{equation}
\label{eq:TCR}
TCR \;=\; \frac{NumOfConfTr}{NumOfTr}.
\end{equation}

where $NumOfConfTr$ denotes the number of valid transactions finalized or confirmed within the observation window, $NumOfTr$ denotes the number of valid transactions submitted within the observation window.

A higher ratio indicates that a larger share of submitted valid transactions was confirmed within the observation window. For example, if $950$ out of $1{,}000$ submitted valid transactions are confirmed under a specified workload and finality criterion, $TCR=0.95$.

\paragraph{Metric Caveats}
Transaction confirmation ratio captures the share of submitted valid transactions confirmed within the observation window. It does not show how long confirmation takes, whether unconfirmed transactions are delayed or dropped, or whether the system remains responsive under different workloads. Results also depend on the finality criterion and on whether faults, partitions, or attacks occur during measurement. Under normal workload conditions, the metric provides scalability evidence; under faults, partitions, or attacks, the same value should be interpreted as security-related evidence about continued operation. Transaction confirmation ratio should therefore be reported together with the workload, observation window, finality criterion, and confirmation-latency measurements.

\constructDefinition{Transaction Throughput}{The number of confirmed transactions processed per time unit under a specified workload or capacity-bound assumption.}

Transaction processing in blockchain systems involves propagating transactions, validating them, batching valid transactions into blocks, and appending those blocks to the main chain. Depending on the consensus/finality model, blocks are either finalized immediately (deterministic finality) or become irreversible after sufficient confirmations (probabilistic finality).

Empirical throughput and capacity-bound throughput capture different quantities: the former measures realized processing under an observed or experimental workload, whereas the latter estimates an upper bound from block-capacity and block-interval parameters under saturated-load assumptions.

The empirical metric computes realized throughput over an observation window, as reported in blockchain benchmarking and measurement studies~(e.g., \cite{ban2019survey, dinh2017blockbench, grabe2020not, gramoli2023diablo, fu2024quantifying, quattrocchi2024blockchain, ren2023bbsf, sanka2021survey, alom2023blockmeter}):

\begin{equation}
\label{eq:throughput-emp}
\mathrm{TPS}_{\mathrm{emp}} \;=\; \frac{\mathrm{NumOfConfTr}}{\,t_1 - t_0\,}\,,
\end{equation}

where $NumOfConfTr$ is the number of transactions that meet the system’s notion of `committed' in $[t_0,t_1]$ (e.g., finalized for immediate finality; $k$-confirmed for probabilistic finality).\footnote{Analysts should align “committed” with the system's finality model and report $k$ where applicable. Ensure $t_1-t_0$ is in seconds to yield TPS.}

A complementary, capacity-bound upper limit follows from per-block constraints and the block interval~\cite{son2021performance}. Let $BCI$ be the block creation interval and
\begin{equation}
\label{eq:NB}
N_B \;=\; \min\!\left(
\left\lfloor \frac{|G_{limit}|}{|G_{cost}|} \right\rfloor,\;
\left\lfloor \frac{B_{\mathrm{byte}}}{s_{\mathrm{tx}}} \right\rfloor
\right),
\end{equation}
be the maximum number of transactions that can fit in a block, where $|G_{limit}|$ is the per-block gas limit (if applicable), $|G_{cost}|$ the average gas cost per transaction, $B_{\mathrm{byte}}$ an optional per-block byte-size limit, and $s_{\mathrm{tx}}$ the average transaction size in bytes. Under saturated load (enough ready transactions at each block boundary), the gas/byte-bound throughput is

\begin{equation}
\label{eq:throughput-cap}
\mathrm{TPS}_{\mathrm{cap}} \;=\; \frac{N_B}{BCI}\,.
\end{equation}

If the arrival rate $\lambda$ of ready transactions is lower than $\mathrm{TPS}_{\mathrm{cap}}$, realized throughput is supply-limited: $\mathrm{TPS} \le \min(\mathrm{TPS}_{\mathrm{cap}}, \lambda)$.
Mempool \emph{capacity} affects queuing and drops, not the per-block packing limit $N_B$.

Realized $\mathrm{TPS}_{\mathrm{emp}}$ is typically below $\mathrm{TPS}_{\mathrm{cap}}$ because of consensus overheads, empty block space, propagation and validation delays, and workload mix.

\paragraph{Metric Caveats}
Empirical and capacity-bound throughput should not be interpreted as the same quantity. $\mathrm{TPS}_{\mathrm{emp}}$ measures realized throughput under a specified workload, observation window, and finality criterion, whereas $\mathrm{TPS}_{\mathrm{cap}}$ estimates an upper bound under saturated load.
Neither captures confirmation latency, transaction drops, non-finalized work, or communication overhead as validating-node count changes.
A $\mathrm{TPS}$ value without the workload, finality rule, observation window, and latency distribution is hard to compare. Two systems can show the same $\mathrm{TPS}$ while giving users very different confirmation experiences.

\subsubsection{Security}
\label{sec:results-sec-constructs}

To estimate security, we identified five constructs that provide partial evidence about whether the blockchain system preserves correct operation and continued progress under faults, partitions, and malicious attacks: \textit{cost of attack}, \textit{fault tolerance}, \textit{reliability}, \textit{\ac{SBR}}, and \textit{success probability of selfish mining}.

\constructDefinition{Cost of Attack}{The estimated fiat-denominated cost required to obtain sufficient consensus-relevant resources to execute a specified attack on a blockchain system.}

Blockchain systems are vulnerable to attacks in which adversaries acquire or coordinate consensus-relevant resources, such as hashing power in \ac{PoW}-based systems or stake in \ac{PoS}-based systems.
A high cost of attack increases the economic barrier to Byzantine behavior, whereas a low cost reduces the barrier to launching attacks.

Equation~\ref{eq:CostofAttack} provides an estimate of the cost of attack $CoA$ for a specified attack vector~\cite{quattrocchi2024blockchain}:
\begin{equation}
\label{eq:CostofAttack}
CoA \;=\; t_h \times c \times \sum_{i=1}^n s_i,
\end{equation}
where $t_h$ denotes the minimum fraction of consensus-relevant resources required to execute the attack, $c$ denotes the monetary cost per unit of that resource, $s_i$ denotes the amount of the resource controlled by validating node $i$, and $n$ denotes the validating-node count. Depending on the blockchain system and attack vector, the relevant resource may be hashing power, stake, voting power, or another protocol-specific resource.

High $CoA$ values indicate that an adversary must incur substantial costs to execute the specified attack, suggesting stronger economic resistance. Conversely, low $CoA$ values indicate that the required resources can be obtained at comparatively low cost, suggesting weaker economic resistance.

This metric applies when attack success can be expressed as control over a quantifiable and economically priced consensus-relevant resource. 
In \ac{PoW}-based systems, this may be hashing power; in \ac{PoS}-based systems, it may be stake or validator voting power.
Such estimates are time-dependent because hardware prices, token prices, total network resources, and liquidity conditions change over time.

\paragraph{Metric Caveats}
$CoA$ provides an economic barrier estimate for a specified attack model. It does not identify the cheapest feasible attack unless multiple attack vectors are compared. Results depend on how effective control over the relevant resource is obtained, since acquisition, rental, and coordination costs vary across attack settings. 
The metric can also omit liquidity constraints, market impact, hardware availability, energy and infrastructure costs, slashing risk, detection risk, loss of attack capital, and attack profitability. It should therefore be interpreted as a cost estimate for a specified attack model, not as a complete measure of security.


\constructDefinition{Fault Tolerance}{The extent to which a blockchain system preserves correct operation and continued progress under a specified fault model.}

Blockchain systems are subject to faults such as crashes, omission, or Byzantine behavior of validating nodes \cite{lamport1982byzantine, fischer1985flp}. Consensus protocols are designed to tolerate some level of faults: for example, Raft handles crash faults, whereas PBFT tolerates Byzantine faults up to a bounded fraction of faulty validating nodes.

Equation~\ref{eq:FT} quantifies fault tolerance as two dimensionless ratios, reported as a pair~\cite{dinh2017blockbench, ren2023bbsf, gramoli2025evaluating}:

\begin{equation}
\label{eq:FT}
\rho_{\mathrm{TPS}}(f) \;=\; \frac{\mathrm{TPS}_{F}(f)}{\mathrm{TPS}_{N}}, \qquad
\rho_{\mathrm{CL}}(f) \;=\; \frac{\mathrm{CL}_{F}(f)}{\mathrm{CL}_{N}},
\end{equation}

where $N$ denotes normal operation and $F$ denotes operation with a share $f$ of faulty validating nodes under a specified fault type, measured with equations~\ref{eq:throughput-emp} and~\ref{eq:emp}. Values of 1 indicate no degradation; $\rho_{\mathrm{TPS}}<1$ and $\rho_{\mathrm{CL}}>1$ indicate degradation, and loss of progress should be reported as such. The largest tolerated $f$ can be compared with the protocol bound (e.g., $f<1/3$ for \ac{BFT}).

\paragraph{Metric Caveats}
This formulation captures performance degradation under specified fault conditions. It does not by itself show whether safety, finality, ledger consistency, or validity is preserved. A blockchain system may maintain throughput and confirmation latency while still violating safety, accepting invalid state transitions, or producing conflicting finalized outcomes. Conversely, a system may show degraded performance while still preserving safety. Fault-tolerance results should therefore report the fault model, workload, baseline condition, faulty condition, measured performance dimensions, and whether the experiment also checks safety- or finality-related outcomes.

\constructDefinition{Reliability}{The probability that the blockchain system continues to provide correct service without failure over a specified time interval.}

Blockchain systems replicate data across validating nodes, often in high-stakes contexts such as finance \cite{kirste2023automated, lamberty2024hybcbdc}. Reliability reflects the probability that the system continues to operate correctly without interruption.

We assume exponentially distributed inter-failure times (homogeneous Poisson failures): a constant hazard rate $\lambda$, so $R(t)=\Pr[T>t]=e^{-\lambda t}$ with $\lambda=1/\mathrm{MTBF}$.

Equation~\ref{eq:reliability} expresses reliability $R(t)$ as a function of the \ac{MTBF}~\cite{grabe2020not}:
\begin{equation}
\label{eq:reliability}
    R(t) \;=\; e^{-\,t/\mathrm{MTBF}}.
\end{equation}

$R(t)$ denotes the probability that the next failure occurs after time $t$.
We estimate $\mathrm{MTBF}$ from observations as
\begin{equation}
\label{eq:MTBF}
    \mathrm{MTBF} = \frac{\mathrm{TotalOprTime}}{\mathrm{NumOfFailures}} .
\end{equation}

where $TotalOprTime$ represents the total time the blockchain system functions correctly within the observation window, $NumOfFailures$ is the count of unique failure events occurring within the observation window.

High reliability indicates that observed failures are rare under the selected failure definition and observation window. Low reliability indicates more frequent observed failures. The implication for security depends on whether the counted failures affect correct operation, continued progress, or both.

If two failures totaling 30 minutes occur in a year (525{,}600 minutes), then
$\mathrm{MTBF} = (525{,}600 - 30)/2 = 262{,}785$ minutes. The probability of experiencing no failure over the full year is $R(525{,}600) = e^{-525{,}600/262{,}785} \approx e^{-2.00} \approx 13.5\%$. Equivalently, there is an $\approx 86.5\%$ probability that at least one failure occurs within that year. If no failures are observed during the observation window, $\mathrm{MTBF}$ should not be computed by division; the result should instead be reported as no observed failure within the observation window, together with the observation duration.

\paragraph{Metric Caveats}
Reliability depends on which failures are counted. Studies may differ in whether they count crashes, stalled finality, inconsistent state, missed slots, unavailable interfaces, or other service interruptions. The exponential model also assumes a constant failure rate; correlated failures, cascading failures, aging effects, or attack-induced failures can violate this assumption. When repair time is non-negligible, reliability should be reported together with the failure definition, observation window, \ac{MTBF}, and \ac{MTTR}.

\constructDefinition{Stale Block Rate}{The ratio of valid propagated blocks that are not included in the canonical chain to confirmed canonical-chain blocks during an observation window.}

In blockchain systems with probabilistic finality, competing valid branches may temporarily exist. When multiple valid blocks are proposed at similar times, only blocks on the eventually selected canonical branch remain part of the main chain, while valid blocks on non-canonical branches become stale blocks~\cite{gervais2016security}. Frequent stale blocks can indicate propagation delays, network partitions, or block-production races, which may facilitate attacks such as double spending or selfish mining~\cite{gervais2016security, sproll2025smsim}.

Equation~\ref{eq:SBR} calculates the \ac{SBR}~\cite{grabe2020not, paulavivcius2021systematic}:
\begin{equation}
\label{eq:SBR}
SBR =  \frac{NumOfStaleBlocks}{|N_c|}
\end{equation}

Here, $NumOfStaleBlocks$ denotes the number of valid blocks proposed but eventually not included in the canonical chain, and $|N_c|$ denotes the number of blocks confirmed on the canonical chain during the observation window.

A higher \ac{SBR} may indicate more frequent temporary forks or chain reorganizations in probabilistic-finality blockchain systems. A lower value suggests fewer observed stale blocks, although exploitability still depends on network conditions and attacker resources.

\paragraph{Metric Caveats}
The metric is simple but backward-looking, measuring past stale blocks without directly predicting attack feasibility. It also neglects system-specific factors, such as block size, block creation interval, propagation delay, and network topology, that influence stale-block occurrence~\cite{gervais2016security}. Moreover, the severity of attacks facilitated by stale blocks is not reflected: in Bitcoin, for instance, rewriting history requires producing a longer chain, which is computationally intensive despite occasional stale blocks~\cite{sproll2025smsim, eyal2018selfish}. Therefore, \ac{SBR} should be interpreted as an observed indicator of fork frequency, not as a complete measure of security.


\constructDefinition{Success Probability of Selfish Mining}{The probability that a selfish validating node's private branch overtakes or manipulates the honest public branch and becomes accepted by the system.}

Selfish mining is an attack strategy in blockchain systems where temporary competing branches can exist. A selfish validating node withholds newly produced blocks to build a private branch and strategically releases it to gain an advantage over honest validating nodes. The attack succeeds when the private branch causes the honest public branch to be displaced or when it increases the attacker's relative rewards compared to honest mining.

Equation~\ref{eq:SPSM} describes how the success probability of selfish mining can be measured~\cite{sproll2025smsim}:

\begin{equation}
\label{eq:SPSM}
P_{\text{success}} =
\frac{N_{succ}}{N_{succ} + N_{fail}}
\end{equation}
where $N_{succ}$ represents the number of selfish mining attempts that succeed, and $N_{fail}$ represents the number of selfish mining attempts that fail.

For illustration, suppose a selfish validating node succeeds in 5 selfish-mining attempts and fails in 10 attempts. The success probability of selfish mining is $5/(5+10)\approx 0.33$.

\paragraph{Metric Caveats}
The metric captures an empirical attack success rate for a specified selfish-mining model, but it does not account for the impact, cost, duration, or profitability of the attack. It also depends on how an attempt and a success are defined. Network conditions, propagation delay, tie-breaking behavior, and the attacker's consensus-relevant resources may affect the observed success probability. Although the same ratio can be adapted to other attack types, such adaptations require attack-specific definitions of success and failure. Therefore, the metric should be interpreted as a selfish-mining-specific probability estimate unless the attack model is explicitly generalized.

\subsection{Applicability Conditions Across Blockchain Systems and Analysis Approaches}
\label{sec:results-applicability-conditions}

Selecting metrics related to the blockchain trilemma requires consideration of the applicability conditions under which these metrics can be used. Tables~\ref{tab:applicability-cond-blockchain} and \ref{tab:applicability-cond-anal} summarize conditions related to blockchain systems and analysis approaches (see Table~\ref{tab:appendix-constructs}). The table should be read as a set of minimum interpretability conditions: an $x$ indicates that the condition must hold for the metric value to support the intended construct interpretation, whereas $(x)$ indicates a condition that is optional or depends on the specific analysis setup.

\begin{table*}[!t]
\renewcommand{\arraystretch}{0.92}
\caption{Overview of applicability conditions associated with blockchain system design.}
\label{tab:applicability-cond-blockchain}
\centering
\footnotesize
\begin{tblr}{
    width=\linewidth,
    rows={valign=m},
    cell{3}{1}={r=4}{valign=m,halign=c},
    cell{7}{1}={r=6}{valign=m,halign=c},
    cell{13}{1}={r=5}{valign=m,halign=c},
    column{1}={0.8cm,c},
    column{2}={2.2cm,l},
    column{3-11}={0.78cm,c},
    column{12}={co=1,halign=l},
    colsep=2.5pt,
    rowsep=1pt,
    row{2}={abovesep=4pt,belowsep=2pt},
    hline{1,Z}={1-12}{0.8pt},
    hline{2}={3-11}{0.5pt},
    hline{3}={1-12}{0.6pt},
    hline{4-6,8-12,14-Y}={1-12}{0.25pt},
    hline{7,13}={1-12}{0.8pt},
    vline{1,Z}={1-Z}{0.8pt},
    vline{2-12}={1-Z}{0.25pt},
    vline{3,12}={1-Z}{0.8pt}
}
\SetCell[r=2]{halign=c,valign=b}\BottomHeader{6.8cm}{\rotatebox{90}{Dimension}}
& \SetCell[r=2]{halign=c,valign=b}\BottomHeader{6.8cm}{Metric}
& \SetCell[c=9]{halign=c}\textbf{Applicability Condition}
& & & & & & & &
& \SetCell[r=2]{halign=c,valign=b}\BottomHeader{6.8cm}{Exemplary Blockchain Systems}
\\
&
& \rotatebox[origin=c]{90}{\begin{minipage}{6.6cm}\raggedright\scriptsize Blockchain system can temporarily exhibit competing valid branches.\end{minipage}}
& \rotatebox[origin=c]{90}{\begin{minipage}{6.6cm}\raggedright\scriptsize Blockchain system relies on probabilistic finality, where blocks may be displaced by competing branches until sufficiently confirmed.\end{minipage}}
& \rotatebox[origin=c]{90}{\begin{minipage}{6.6cm}\raggedright\scriptsize Blockchain system produces capacity-constrained blocks, slots, or batches at a known or configurable interval.\end{minipage}}
& \rotatebox[origin=c]{90}{\begin{minipage}{6.6cm}\raggedright\scriptsize Blockchain system uses epoch- or slot-based finality with known finality parameters.\end{minipage}}
& \rotatebox[origin=c]{90}{\begin{minipage}{6.6cm}\raggedright\scriptsize Consensus influence can be quantified per validating node, e.g., hashing power, stake, or voting weight.\end{minipage}}
& \rotatebox[origin=c]{90}{\begin{minipage}{6.6cm}\raggedright\scriptsize Consensus-relevant resources required for an attack can be quantified and priced.\end{minipage}}
& \rotatebox[origin=c]{90}{\begin{minipage}{6.6cm}\raggedright\scriptsize Number of subsequent blocks increases the security of previous blocks added to the blockchain system.\end{minipage}}
& \rotatebox[origin=c]{90}{\begin{minipage}{6.6cm}\raggedright\scriptsize Protocol-native token amounts relevant to consensus influence can be attributed to validating nodes or controlling entities.\end{minipage}}
& \rotatebox[origin=c]{90}{\begin{minipage}{6.6cm}\raggedright\scriptsize Validating nodes are selected (pseudo-)randomly for block proposal.\end{minipage}}
& \\

\rotatebox[origin=c]{90}{\textbf{DoD}}
& $H$ (eq.~\ref{eq:BPR}) & & & & & & & & & x & Bitcoin, Ethereum, Tezos \\
& $NC$ (eq.~\ref{eq:CRD}) & & & & & x & & & & & Bitcoin \\
& $HHI$ (eq.~\ref{eq:TC}) & & & & & & & & x & & Bitcoin, Ethereum, Tezos \\
& $Gini$ (eq.~\ref{eq:WD_Gini}) & & & & & & & & x & & Bitcoin, Ethereum, Tezos \\

\rotatebox[origin=c]{90}{\textbf{Scalability}}
& $CL_{conf}$ (eq.~\ref{eq:conf}) & x & x & & & & & x & & & Bitcoin \\
& $CL_{emp}$ (eq.~\ref{eq:emp}) & & & & & & & x & & & Bitcoin, Ethereum, Hyperledger Fabric, Algorand, Solana \\
& $CL_{epoch}$ (eq.~\ref{eq:epoch}) & & & & x & & & & & & Ethereum \\
& $TCR$ (eq.~\ref{eq:TCR}) & x & x & & & & & x & & & Algorand, Avalanche, Ethereum, Hyperledger Fabric, \newline Red Belly, Solana \\
& $TPS_{cap}$ (eq.~\ref{eq:throughput-cap}) & & & x & & & & & & & Bitcoin, Ethereum, Polygon PoS, Binance Smart Chain, Avalanche C-Chain, Fantom, Tezos\\
& $\mathrm{TPS}_{\mathrm{emp}}$ (eq.~\ref{eq:throughput-emp}) & & & & & & & & & & Algorand, Avalanche, Binance Smart Chain, Bitcoin, Cardano, Elrond, Ethereum, Fantom, Hyperledger Fabric, Polygon, Quorum, Red Belly, Solana, Tezos, Tron \\

\rotatebox[origin=c]{90}{\textbf{Security}}
& $CoA$ (eq.~\ref{eq:CostofAttack}) & & & & & x & x & & & & Bitcoin, Ethereum \\
& $FT$ (eq.~\ref{eq:FT}) & & & & & & & & & & Algorand, BitShares, Ethereum, Hyperledger Burrow, Hyperledger Fabric, Quorum, Stellar \\
& $R(t)$ (eq.~\ref{eq:reliability}) & & & & & & & & & & Ethereum, Hyperledger Indy, Tezos \\
& $SBR$ (eq.~\ref{eq:SBR}) & x & x & & & & & & & & Bitcoin \\
& $SPSM$ (eq.~\ref{eq:SPSM}) & x & x & & & & & & & & Bitcoin \\
\end{tblr}

\vspace{2mm}
\begin{minipage}{\linewidth}
\scriptsize
$x$ = required applicability condition;
$(x)$ = optional applicability condition;
eq. = equation;
$H$ = Shannon entropy;
$NC$ = Nakamoto coefficient;
$HHI$ = Herfindahl--Hirschman Index;
$Gini$ = Gini coefficient;
$CL_{conf}$ = Confirmation-depth latency;
$CL_{emp}$ = Empirical confirmation latency;
$CL_{epoch}$ = Epoch-based confirmation latency;
$TCR$ = Transaction confirmation ratio;
$TPS_{cap}$ = Capacity-bound transaction throughput;
$\mathrm{TPS}_{\mathrm{emp}}$ = Empirical transaction throughput;
$CoA$ = Cost of attack;
$FT$ = Fault tolerance;
$R(t)$ = Reliability;
$SBR$ = Stale block rate;
$SPSM$ = Success probability of selfish mining.
\end{minipage}
\end{table*}

\begin{table*}[!t]
\renewcommand{\arraystretch}{0.92}
\caption{Overview of applicability conditions associated with analysis approaches.}
\label{tab:applicability-cond-anal}
\centering
\footnotesize
\begin{tblr}{
    width=\linewidth,
    rows={valign=m},
    cell{3}{1}={r=4}{valign=m,halign=c},
    cell{7}{1}={r=6}{valign=m,halign=c},
    cell{13}{1}={r=5}{valign=m,halign=c},
    column{1}={0.8cm,c},
    column{2}={2.2cm,l},
    column{3-12}={0.78cm,c},
    column{13}={co=1,halign=l},
    colsep=2.5pt,
    rowsep=1pt,
    row{2}={abovesep=4pt,belowsep=2pt},
    hline{1,Z}={1-13}{0.8pt},
    hline{2}={3-12}{0.5pt},
    hline{3}={1-13}{0.6pt},
    hline{4-6,8-12,14-17}={1-13}{0.25pt},
    hline{7,13}={1-13}{0.8pt},
    vline{1,Z}={1-Z}{0.8pt},
    vline{2-13}={1-Z}{0.25pt},
    vline{3,13}={1-Z}{0.8pt}
}
\SetCell[r=2]{halign=c,valign=b}\BottomHeader{6.8cm}{\rotatebox{90}{Dimension}}
& \SetCell[r=2]{halign=c,valign=b}\BottomHeader{6.8cm}{Metric}
& \SetCell[c=10]{halign=c}\textbf{Applicability Condition}
& & & & & & & & &
& \SetCell[r=2]{halign=c,valign=b}\BottomHeader{6.8cm}{Exemplary Blockchain Systems}
\\
&
& \rotatebox[origin=c]{90}{\begin{minipage}{6.6cm}\raggedright\scriptsize Attack success and failure criteria are defined.\end{minipage}}
& \rotatebox[origin=c]{90}{\begin{minipage}{6.6cm}\raggedright\scriptsize Attribution level is defined.\end{minipage}}
& \rotatebox[origin=c]{90}{\begin{minipage}{6.6cm}\raggedright\scriptsize Block capacity and block interval parameters are known.\end{minipage}}
& \rotatebox[origin=c]{90}{\begin{minipage}{6.6cm}\raggedright\scriptsize Block creation interval and required confirmation depth are known.\end{minipage}}
& \rotatebox[origin=c]{90}{\begin{minipage}{6.6cm}\raggedright\scriptsize Block-proposal and confirmation/finalization timestamps are observable.\end{minipage}}
& \rotatebox[origin=c]{90}{\begin{minipage}{6.6cm}\raggedright\scriptsize Block-proposal outcomes are observable.\end{minipage}}
& \rotatebox[origin=c]{90}{\begin{minipage}{6.6cm}\raggedright\scriptsize Confirmation/finalization criteria are specified.\end{minipage}}
& \rotatebox[origin=c]{90}{\begin{minipage}{6.6cm}\raggedright\scriptsize Epoch/slot finality parameters are known.\end{minipage}}
& \rotatebox[origin=c]{90}{\begin{minipage}{6.6cm}\raggedright\scriptsize Failure events are observable and countable.\end{minipage}}
& \rotatebox[origin=c]{90}{\begin{minipage}{6.6cm}\raggedright\scriptsize Failure model is specified.\end{minipage}}
&
\\

\rotatebox[origin=c]{90}{\textbf{DoD}}
& $H$ (eq.~\ref{eq:BPR}) & & & & & & x & & & & & Bitcoin, Ethereum, Tezos \\
& $NC$ (eq.~\ref{eq:CRD}) & & x & & & & & & & & & Bitcoin \\
& $HHI$ (eq.~\ref{eq:TC}) & & x & & & & & & & & & Bitcoin, Ethereum, Tezos \\
& $Gini$ (eq.~\ref{eq:WD_Gini}) & & x & & & & & & & & & Bitcoin, Ethereum, Tezos \\

\rotatebox[origin=c]{90}{\textbf{Scalability}}
& $CL_{conf}$ (eq.~\ref{eq:conf}) & & & & x & & & & & & & Bitcoin \\
& $CL_{emp}$ (eq.~\ref{eq:emp}) & & & & & & & x & & & & Bitcoin, Ethereum, Hyperledger Fabric, Algorand, Solana \\
& $CL_{epoch}$ (eq.~\ref{eq:epoch}) & & & & & & x & & x & & & Ethereum \\
& $TCR$ (eq.~\ref{eq:TCR}) & & & & & & & x & & & & Algorand, Avalanche, Ethereum, Hyperledger Fabric, \newline Red Belly, Solana \\
& $TPS_{cap}$ (eq.~\ref{eq:throughput-cap}) & & & x & & & & & & & & \\
& $\mathrm{TPS}_{\mathrm{emp}}$ (eq.~\ref{eq:throughput-emp}) & & & & & & & & & & & Algorand, Avalanche, Binance Smart Chain, Bitcoin, Cardano, Elrond, Ethereum, Fantom, Hyperledger Fabric, Polygon, Quorum, Red Belly, Solana, Tezos, Tron. \\
\rotatebox[origin=c]{90}{\textbf{Security}}
& $CoA$ (eq.~\ref{eq:CostofAttack}) & & & & & & & & & & & Bitcoin, Ethereum \\
& $FT$ (eq.~\ref{eq:FT}) & & & & & & & & & & & Algorand, BitShares, Ethereum, Hyperledger Burrow, Hyperledger Fabric, Quorum, Stellar \\
& $R(t)$ (eq.~\ref{eq:reliability}) & & & & & & & & & x & x & Ethereum, Hyperledger Indy, Tezos \\
& $SBR$ (eq.~\ref{eq:SBR}) & & & & & & & & & & & Bitcoin \\
& $SPSM$ (eq.~\ref{eq:SPSM}) & x & & & & & & & & & & Bitcoin \\
\end{tblr}
\end{table*}

\begin{table*}[!t]
\renewcommand{\arraystretch}{0.92}
\addtocounter{table}{-1} \caption{(Continued) Overview of applicability conditions associated with analysis approaches.}
\centering
\footnotesize
\begin{tblr}{
    width=\linewidth,
    rows={valign=m},
    cell{3}{1}={r=4}{valign=m,halign=c},
    cell{7}{1}={r=6}{valign=m,halign=c},
    cell{13}{1}={r=5}{valign=m,halign=c},
    column{1}={0.8cm,c},
    column{2}={2.2cm,l},
    column{3-11}={0.78cm,c},
    column{12}={co=1,halign=l},
    colsep=2.5pt,
    rowsep=1pt,
    row{2}={abovesep=4pt,belowsep=2pt},
    hline{1,Z}={1-12}{0.8pt},
    hline{2}={3-11}{0.5pt},
    hline{3}={1-12}{0.6pt},
    hline{4-6,8-12,14-17}={1-12}{0.25pt},
    hline{7,13}={1-12}{0.8pt},
    vline{1,Z}={1-Z}{0.8pt},
    vline{2-12}={1-Z}{0.25pt},
    vline{3,12}={1-Z}{0.8pt}
}
\SetCell[r=2]{halign=c,valign=b}\BottomHeader{6.8cm}{\rotatebox{90}{Dimension}}
& \SetCell[r=2]{halign=c,valign=b}\BottomHeader{6.8cm}{Metric}
& \SetCell[c=9]{halign=c}\textbf{Applicability Condition}
& & & & & & & &
& \SetCell[r=2]{halign=c,valign=b}\BottomHeader{6.8cm}{Exemplary Blockchain Systems}
\\
&
& \rotatebox[origin=c]{90}{\begin{minipage}{6.6cm}\raggedright\scriptsize Fault conditions can be observed or induced.\end{minipage}}
& \rotatebox[origin=c]{90}{\begin{minipage}{6.6cm}\raggedright\scriptsize Fault-free and faulty operating conditions can be compared.\end{minipage}}
& \rotatebox[origin=c]{90}{\begin{minipage}{6.6cm}\raggedright\scriptsize Observation window is defined and comparable.\end{minipage}}
& \rotatebox[origin=c]{90}{\begin{minipage}{6.6cm}\raggedright\scriptsize Relevant attack threshold is specified.\end{minipage}}
& \rotatebox[origin=c]{90}{\begin{minipage}{6.6cm}\raggedright\scriptsize Stale and confirmed blocks are observable over an observation window.\end{minipage}}
& \rotatebox[origin=c]{90}{\begin{minipage}{6.6cm}\raggedright\scriptsize Successful and failed attack attempts can be observed or simulated.\end{minipage}}
& \rotatebox[origin=c]{90}{\begin{minipage}{6.6cm}\raggedright\scriptsize Transaction outcomes are observable.\end{minipage}}
& \rotatebox[origin=c]{90}{\begin{minipage}{6.6cm}\raggedright\scriptsize Transaction workload can be controlled or characterized.\end{minipage}}
& \rotatebox[origin=c]{90}{\begin{minipage}{6.6cm}\raggedright\scriptsize Validating-node counts are observable.\end{minipage}}
&
\\

\rotatebox[origin=c]{90}{\textbf{DoD}}
& $H$ (eq.~\ref{eq:BPR}) & & & & x & & & & & x & Bitcoin, Ethereum, Tezos \\
& $NC$ (eq.~\ref{eq:CRD}) & & & & & x & & & & x & Bitcoin \\
& $HHI$ (eq.~\ref{eq:TC}) & & & & & & & & & x & Bitcoin, Ethereum, Tezos \\
& $Gini$ (eq.~\ref{eq:WD_Gini}) & & & & & & & & & x & Bitcoin, Ethereum, Tezos \\

\rotatebox[origin=c]{90}{\textbf{Scalability}}
& $CL_{conf}$ (eq.~\ref{eq:conf}) & & x & & & & & & & & Bitcoin \\
& $CL_{emp}$ (eq.~\ref{eq:emp}) & & & (x) & & & & x & (x) & & Bitcoin, Ethereum, Hyperledger Fabric, Algorand, Solana \\
& $CL_{epoch}$ (eq.~\ref{eq:epoch}) & & & & & & & & & & Ethereum \\
& $TCR$ (eq.~\ref{eq:TCR}) & & & x & & & & x & x & & Algorand, Avalanche, Ethereum, Hyperledger Fabric, \newline Red Belly, Solana \\
& $TPS_{cap}$ (eq.~\ref{eq:throughput-cap}) & & & & & & & & & & \\
& $\mathrm{TPS}_{\mathrm{emp}}$ (eq.~\ref{eq:throughput-emp}) & & x & & & x & & & & x & Algorand, Avalanche, Binance Smart Chain, Bitcoin, Cardano, Elrond, Ethereum, Fantom, Hyperledger Fabric, Polygon, Quorum, Red Belly, Solana, Tezos, Tron. \\
\rotatebox[origin=c]{90}{\textbf{Security}}
& $CoA$ (eq.~\ref{eq:CostofAttack}) & & & & x & & & & & (x) & Bitcoin, Ethereum \\
& $FT$ (eq.~\ref{eq:FT}) & (x) & x & x & & & & & x & & Algorand, BitShares, Ethereum, Hyperledger Burrow, Hyperledger Fabric, Quorum, Stellar \\
& $R(t)$ (eq.~\ref{eq:reliability}) & & & x & & & & & & & Ethereum, Hyperledger Indy, Tezos \\
& $SBR$ (eq.~\ref{eq:SBR}) & & x & x & & & & & & & Bitcoin \\
& $SPSM$ (eq.~\ref{eq:SPSM}) & & (x) & x & & & x & & & & Bitcoin \\
\end{tblr}
\end{table*}

For \ac{DoD}, metrics capture different forms of influence on consensus of validating nodes. Consensus-resource and token-distribution metrics are suitable when the selected resource or token amount can be quantified, attributed, and linked to influence on consensus.

Scalability metrics apply across a broader range of blockchain systems, but their interpretation depends on workload characteristics, transaction outcomes, finality criteria, and observation windows.
Transaction confirmation ratio requires special care: in this work, it is a scalability construct only when it measures the share of submitted valid transactions confirmed under a specified workload.
If the same ratio is measured under crash faults, Byzantine behavior, attacks, or network partitions, it should instead be interpreted as security-related evidence about continued operation.

Security metrics depend on the specified fault model, attack model, resource assumptions, and observation window. Cost of attack applies when attack success can be expressed as control over a quantifiable and economically priced consensus-relevant resource. Fault tolerance applies when fault-free and faulty operating conditions can be compared under a specified fault model. Stale block rate and success probability of selfish mining are most informative in blockchain systems where temporary competing branches can occur.

    \section{Discussion}
\label{sec:discussion}

The main difficulty in trilemma-oriented analysis is not the absence of metrics, but the interpretation of what a metric value can support. A metric value becomes meaningful for the blockchain trilemma only when the measured object is linked to a construct, the construct is linked to a trilemma dimension, and the system assumptions under which this interpretation holds are made explicit.
In this section, we discuss the principal findings, contributions, limitations, and open research challenges that follow from this interpretation problem.

\subsection{Principal Findings}
\label{sec:principal-findings}

Our synthesis exposes a problem that is easily hidden in trilemma-oriented analysis: selecting a metric is not sufficient unless the interpretation from metric to construct and from construct to dimension is made explicit. Prior work provides benchmark metrics, decentralization measurements, trilemma evaluation frameworks, and formal or empirical analyses of selected aspects of \ac{DoD}, scalability, and security \cite{dinh2017blockbench, gramoli2023diablo, ren2023bbsf, zhang2022sok, ovezik2025sok, nakai2024formulation, quattrocchi2024blockchain}. Yet this literature often leaves implicit how a measured value supports the construct it operationalizes and how this construct supports an interpretation of a trilemma dimension. The constructs identified in this study should therefore not be read as alternative measures of entire dimensions. They capture specific aspects of dimensions that remain broad and internally heterogeneous. The caveats reported in Section~\ref{sec:results-constructs} show where this interpretation can break down: a metric may be correctly computed while still providing weak evidence for \ac{DoD}, scalability, or security if the measured object is only weakly connected to influence on consensus, workload, and finality assumptions, or the relevant fault and attack model.

This problem is easier to delimit for scalability than for \ac{DoD} and security. Constructs such as \emph{transaction throughput} and \emph{confirmation latency} build on established performance concepts in blockchain benchmarking, although their interpretation still depends on workload and finality criteria \cite{dinh2017blockbench, gramoli2023diablo, ren2023bbsf}. For \ac{DoD}, the connection is less direct. A distribution of resources, tokens, or block proposals is not already a distribution of equitable influence on consensus.
It supports an interpretation of \ac{DoD} only when the measured resource or observed task actually shapes consensus-relevant influence, and when the attribution level reflects the validating nodes or entities that exercise this influence \cite{zhang2022sok, ovezik2025sok, muellerbloch2024dodpos}. Security is similarly difficult to compress into a single construct because prior work often analyzes specific fault assumptions, attack strategies, or continued operation under selected conditions rather than security as a whole \cite{gervais2016security, bonneau2016buy, karame2012double}. Aggregate scores for \ac{DoD}, scalability, or security would require explicit assumptions about overlap, substitutability, and compensation among constructs. Without such assumptions, aggregation risks turning partial evidence from constructs into a score that appears more precise than the underlying interpretation allows.

This also changes how trade-off expectations should be formulated. The blockchain trilemma is useful because it frames recurring tensions among \ac{DoD}, scalability, and security \cite{kannengiesser2020trade, xiao2020survey, werth2023review, nakai2024formulation, mssassi2024blockchain, quattrocchi2024blockchain}. However, our synthesis suggests that these tensions are not stable relationships among dimensions. They emerge through design mechanisms that connect particular constructs. Assigning ordering or proposal influence to fewer validating nodes can improve \emph{transaction throughput} or \emph{confirmation latency}, but it reduces \ac{DoD} only if the influence of validating nodes on consensus becomes less equitable. Higher transaction throughput affects security only if the mechanism that produces it weakens propagation, validation, finality, or fault assumptions under which the core blockchain system preserves correct operation and continued progress. Conversely, a broader distribution of consensus-relevant resources can support both \ac{DoD} and security because an adversary must coordinate or control more consensus-relevant influence to affect correct operation or continued progress \cite{kannengiesser2020trade, gervais2016security, bonneau2016buy}.
The blockchain trilemma is therefore most defensible as a mechanism-level analytical lens: a trade-off claim should specify which constructs are linked, which design mechanism links them, and under which system and analysis assumptions the relationship is expected to hold.

\subsection{Contributions}
\label{sec:contributions-to-research-and-practice}

This study contributes a measurement-oriented account of the blockchain trilemma. Rather than treating \ac{DoD}, scalability, and security as directly measurable dimensions, it shows how trilemma-oriented analysis depends on an intermediate layer of constructs that specify what aspect of a dimension is actually being observed. This distinction is important because the same metric can support different conclusions depending on the construct it is used to operationalize. The construct--metric synthesis therefore provides a more defensible basis for metric selection: researchers can state more precisely what their measurements represent, and practitioners can select metrics that correspond to the quality concerns of a particular blockchain system design.

The study also contributes a set of interpretation conditions for trilemma-oriented measurement. Metric values are not self-explanatory evidence for decentralization, scalability, or security. Their meaning depends on system-specific assumptions, including role structure, consensus-relevant resources, finality model, workload, and fault or attack model. By making these conditions explicit, the synthesis helps avoid a common problem in blockchain-system evaluation: reporting a metric value as if it measured an entire trilemma dimension, although it only captures a narrower and design-dependent construct.

Finally, the study links metric interpretation to the data and analysis approach required to compute the metric.
The distinction between design input parameters and environmental input parameters clarifies which factors belong to the blockchain system design and which belong to the measurement context. This distinction is relevant for benchmark design, simulation studies, and analytical comparisons because it makes clear which parameters can be varied as architectural choices and which must be controlled, observed, or reported as part of the evaluation setting. Together with the mapping to analytical, simulation-based, and benchmarking approaches (see Appendix~\ref{appdx:ovw-analysis-approaches}), this supports more transparent and comparable empirical analysis of blockchain systems through the lens of the blockchain trilemma.

\subsection{Limitations}
\label{sec:limitations}

The scope of this study is conceptual harmonization and verification rather than empirical validation. The study verifies operationalizations by assessing whether construct--metric combinations are internally consistent, traceable to prior literature, specified clearly enough to be computed, and applicable under stated system assumptions. This differs from validating dimension-level interpretations. Such validation would require showing that measured construct-level evidence supports claims about \ac{DoD}, scalability, or security in real blockchain system designs. This study does not provide such validation. Its results should therefore be read as a practical basis for trilemma-oriented analysis, not as evidence that the identified construct--metric combinations are sufficient for all systems, architectures, or analysis objectives.

The literature corpus may also reflect sampling constraints. Although the systematic database search supports rigor, the supplementary Google Scholar search was purposive rather than exhaustive and was used to improve conceptual coverage. We did not record a complete count of all Google Scholar results screened. The corpus may underrepresent recent preprints, system documentation, non-English venues, white papers, or implementation changes that have not undergone peer review. As a result, some emerging constructs, metrics, or applicability conditions may not yet be represented in the reviewed literature.

Finally, applicability conditions were synthesized from information explicitly reported in the literature and from characteristics inferred from metric definitions, blockchain system properties, and experimental descriptions. Such interpretations involve researcher judgment despite iterative coding and author feedback. Some applicability conditions may therefore be refined as additional empirical evidence becomes available or as blockchain system architectures evolve.

\subsection{Open Research Challenges}
\label{sec:future-work}

Several open challenges remain for future research.

\paragraph{From Construct--Metric Selection to Dimension Interpretation.}
This study provides a foundation for selecting construct--metric combinations for \ac{DoD}, scalability, and security. Future research can build on this foundation by examining how several constructs within the same dimension jointly support a more expressive interpretation of that dimension. This requires explicit assumptions about whether constructs overlap, complement each other, or capture distinct aspects that cannot compensate for one another. Such work is particularly relevant for \ac{DoD} and security, where no single construct captures the dimension as a whole.

\paragraph{Contextualizing the Blockchain Trilemma.}
The blockchain trilemma is abstract and becomes more analytically useful when it is contextualized into concrete blockchain system designs. Future research should therefore specify context-specific trilemmas \cite{hong2014framework}, for example by making explicit the validating-node roles, finality model, and fault or attack model under consideration. Within such contextualized trilemmas, researchers can formulate hypotheses about which constructs are linked by which design mechanisms and quantify these relationships through suitable construct--metric combinations. This would help explain not only whether expected trade-offs occur, but also how their observed strength and direction depend on the selected constructs, metrics, and caveats.

\paragraph{Layered Blockchain Architectures.}
Layer-2 systems, sidechains, and state channels do not resolve the trilemma within the core blockchain-system boundary used in this study. They add mechanisms that redistribute functions and dependencies across a stack of layers, including execution, settlement, data availability, dispute resolution, and ordering roles. Future research should therefore first develop system models that make Layer-1/Layer-2 dependencies explicit before adapting construct--metric combinations. Such models would need to specify which properties are inherited from the core blockchain system, which are introduced by the additional layer, and where ordering-, settlement-, or consensus-relevant influence is located.

\paragraph{Sociotechnical Extensions of the System Model.}
The constructs synthesized in this study primarily capture technical or economic aspects of blockchain system designs. From an information-systems perspective, future research could extend the system model to represent social aspects of the blockchain-based information systems, including relationships among validating-node controllers, governance structures, and organizational affiliations \cite{chatterjee2021possible, kannengiesser2020trade}. Such extensions are relevant when these aspects affect equitable influence on consensus or the autonomous exercise of this influence, for example through coordination or collusion that is not visible in node-level or resource-level metrics.
Related work should also examine whether transaction-ordering power, proposer-builder separation, relay markets, or \ac{MEV} affect the interpretation of \ac{DoD} or security in blockchain systems.
    \section{Conclusion}
\label{sec:conclusion}

The blockchain trilemma is most useful when treated as an analytical lens rather than as a directly measurable object. Meaningful analysis therefore requires an explicit link from a trilemma dimension to the construct being analyzed and from that construct to the metric used to operationalize it. Without this link, correctly computed metric values can still support misleading conclusions about \ac{DoD}, scalability, or security.

Measurement for trilemma-oriented analysis is conditional on blockchain system design. Claims about trade-offs should therefore specify the measured construct, the metric, and the system assumptions under which the value is interpreted. This shifts the use of the blockchain trilemma from a broad design intuition toward a more explicit basis for comparing blockchain system designs and reasoning about design trade-offs.

	\bibliographystyle{IEEEtran}
	\bibliography{references}

    \appendices
     \balance
     \enlargethispage{2\baselineskip}
\section*{Supplementary Material}
\setcounter{table}{0}
\renewcommand{\thetable}{A\arabic{table}}

\setcounter{subsection}{0}
\renewcommand{\thesubsection}{A\arabic{subsection}}
\renewcommand{\thesubsectiondis}{A\arabic{subsection}.}

\subsection{Overview of Input Parameters}
\label{appdx:ovw-input-parameters}

Table~\ref{tab:input-parameters} summarizes the input parameters used in these metrics and classifies them as design or environmental. Design input parameters are fixed by protocol design, system configuration, or the selected analysis criterion. Environmental input parameters are observed from the analyzed deployment, workload, resource distribution, market condition, or experiment execution.

\begin{table*}[!t]
\renewcommand{\arraystretch}{1.15}
\caption{Overview of input parameters of metrics that operationalize constructs associated with the blockchain trilemma dimensions.}
\label{tab:input-parameters}
\centering
\footnotesize
\begin{tblr}{
    width = \textwidth,
    cell{2}{1} = {r=23}{valign=m,halign=c},
    column{1} = {0.7cm},
    column{2} = {3.6cm},
    column{3} = {2.8cm,halign=c},
    column{4} = {co=1},
    column{5} = {1.9cm,halign=c},
    row{1} = {font=\bfseries,halign=c},
    rows = {valign=m},
    hline{1} = {2-5}{},
    hline{2-Z} = {1-5}{},
    vline{1} = {2-Z}{},
    vline{2-6} = {1-Z}{}
}

& \textbf{Input Parameter} & \textbf{Symbol} & \textbf{Description} & \textbf{Equations} \\

\rotatebox[origin=c]{90}{\textbf{Environmental}}
& Accumulated Resources
& $s_i$
& The amount of the economically priced consensus-relevant resource attributed to validating node or analysis unit $i$.
& \ref{eq:CostofAttack} \\

& Attribution Unit Count
& $m$
& The number of observed attribution units over which consensus-relevant resource shares are measured.
& \ref{eq:CRD} \\

& Confirmation Time
& $ConfirmTime$
& The timestamp at which a submitted transaction or proposed block satisfies the reported confirmation or finalization criterion.
& \ref{eq:emp} \\

& Consensus-Relevant Resource Share
& $p$
& The proportion of a consensus-relevant resource controlled by an entity, such as hashing power in \ac{PoW} systems, validating node stake in \ac{PoS} systems, delegated voting power, or other protocol-defined measures of influence.
& \ref{eq:CRD} \\

& Consensus-Relevant Token Amount
& $\tau_i$
& The amount of protocol-native tokens attributed to validating node $i$ that is relevant to consensus participation, influence on consensus, or reward accumulation.
& \ref{eq:TC}, \ref{eq:WD_Gini} \\

& Elapsed Time
& $t$
& The time horizon over which reliability is estimated.
& \ref{eq:reliability} \\

& Number of Confirmed Blocks
& $|N_c|$
& The number of blocks confirmed on the canonical chain during the observation window.
& \ref{eq:SBR} \\

& Number of Confirmed Transactions
& $NumOfConfTr$
& The total number of submitted valid transactions that satisfy the reported confirmation or finalization criterion during the observation window.
& \ref{eq:TCR}, \ref{eq:throughput-emp} \\

& Number of Failed Selfish Mining Attempts
& $N_{fail}$
& The number of selfish mining attempts that do not achieve the intended outcome and therefore fail to provide the expected advantage or reward to the attacker.
& \ref{eq:SPSM} \\

& Number of Failures
& $NumOfFailures$
& The count of distinct failure events occurring in a blockchain system within a given observation period.
& \ref{eq:MTBF} \\

& Number of Proposed Blocks
& $b_i$
& The number of blocks proposed by validating nodes to a blockchain system.
& \ref{eq:BPR} \\

& Number of Stale Blocks
& $NumOfStaleBlocks$
& The number of valid propagated blocks that are not included in the canonical chain during the observation window.
& \ref{eq:SBR} \\

& Number of Submitted Valid Transactions
& $NumOfTr$
& The total number of valid transactions submitted to the blockchain system during the observation window.
& \ref{eq:TCR} \\

& Number of Successful Selfish Mining Attempts
& $N_{succ}$
& The number of selfish mining attempts that successfully result in the attacker achieving the intended outcome.
& \ref{eq:SPSM} \\

& Observation Window End
& $t_1$
& The end of the observation window used to estimate empirical transaction throughput.
& \ref{eq:throughput-emp} \\

& Observation Window Start
& $t_0$
& The start of the observation window used to estimate empirical transaction throughput.
& \ref{eq:throughput-emp} \\

& Resource Cost
& $c$
& The monetary cost per unit of resources, such as token or hashing power.
& \ref{eq:CostofAttack} \\

& Start Time
& $StartTime$
& The timestamp from which confirmation latency is measured, such as transaction-submission time in transaction-level benchmarks or block-proposal time in block-level analyses.
& \ref{eq:emp} \\

& Total Consensus-Relevant Token Amount
& $\tau_{total}$
& The total consensus-relevant token amount across all validating nodes in the observation window.
& \ref{eq:TC} \\

& Total Operational Time
& $TotalOprTime$
& The timespan a blockchain system operates correctly within a defined observation time.
& \ref{eq:reliability} \\

& Transaction Gas Cost
& $\left|G_{cost}\right|$
& The average gas required by the transaction type or workload used for the capacity-bound throughput estimate.
& \ref{eq:NB} \\

& Transaction Size
& $s_{\mathrm{tx}}$
& The average size of a blockchain transaction in bytes.
& \ref{eq:NB} \\

& Validating-Node Count
& $n$
& The observed number of validating nodes included in the metric computation.
& \ref{eq:BPR}, \ref{eq:TC}, \ref{eq:CostofAttack}, \ref{eq:WD_Gini} \\
\end{tblr}
\end{table*}

\begin{table*}[!t]
\renewcommand{\arraystretch}{1.15}
\addtocounter{table}{-1}
\caption{(Continued) Overview of input parameters of metrics that operationalize constructs associated with the blockchain trilemma dimensions.}
\centering
\footnotesize
\begin{tblr}{
    width = \textwidth,
    cell{2}{1} = {r=9}{valign=m,halign=c},
    column{1} = {0.7cm},
    column{2} = {3.6cm},
    column{3} = {2.8cm,halign=c},
    column{4} = {co=1},
    column{5} = {1.9cm,halign=c},
    row{1} = {font=\bfseries,halign=c},
    rows = {valign=m},
    hline{1} = {2-5}{},
    hline{2-Z} = {1-5}{},
    vline{1} = {2-Z}{},
    vline{2-6} = {1-Z}{}
}

& \textbf{Input Parameter} & \textbf{Symbol} & \textbf{Description} & \textbf{Equations} \\

\rotatebox[origin=c]{90}{\textbf{Design}}
& Attack Resource Threshold
& $t_h$
& The minimum fraction of the economically priced consensus-relevant resource required to execute the specified attack.
& \ref{eq:CostofAttack} \\

& Block Byte Limit
& $B_{\mathrm{byte}}$
& The maximum byte-size capacity available for transactions in a block, if such a limit is used by the blockchain system.
& \ref{eq:NB} \\

& Block Creation Interval
& $BCI$
& The average time in milliseconds between the proposal of consecutive blocks that are included in a blockchain.
& \ref{eq:conf}, \ref{eq:NB} \\

& Block Gas Limit
& $\left|G_{limit}\right|$
& The maximum amount of computational resources available per block, if the blockchain system uses a gas-based block-capacity limit.
& \ref{eq:NB} \\

& Compromise Threshold
& $\theta$
& The threshold share of a consensus-relevant resource required to exceed the selected compromise condition for the Nakamoto coefficient.
& \ref{eq:CRD} \\

& Confirmation Depth
& $b^\prime$
& The selected number of subsequent canonical-chain blocks required before a block is treated as sufficiently confirmed.
& \ref{eq:conf} \\

& Epoch Length
& $e_l$
& The fixed number of slots that constitute a single epoch in a blockchain system.
& \ref{eq:TTJ} \\

& Epoch to Finality
& $e_f$
& The number of epochs required for a block to be considered finalized in a blockchain system.
& \ref{eq:epoch} \\

& Slot Time
& $slot_{\mathit{time}}$
& The duration of a slot in an epoch- or slot-based blockchain system.
& \ref{eq:TTJ} \\
\end{tblr}

\vspace{4mm}

\centering
\scriptsize
\caption{Overview of analysis approaches for investigating the blockchain trilemma dimensions with examples of blockchain systems analyzed using these approaches.}
\label{tab:appendix-constructs}
\begin{tblr}{
    width = \textwidth,
    hspan = minimal,
    colspec = {Q[2.9cm,l] Q[2.1cm,c] Q[1.9cm,c] Q[1.6cm,c] X[l]},
    rows = {valign = m},
    row{1-2} = {font = \bfseries},
    colsep = 3pt,
    rowsep = 0.6pt,
    hline{1} = {2-4}{},
    hline{2-Z} = {1-Z}{},
    vline{2,5} = {1}{},
    vline{1-Z} = {2-Z}{}
}
 & \SetCell[c=3]{c} Operationalized Constructs Used to Estimate the Blockchain Trilemma Dimensions & & & \\
Analysis Approach & DoD & Scalability & Security & Analyzed Blockchain Systems \\
AlgorandStressTester \cite{esmaili2025experimental} & & CL, TT & & Algorand \\
Ban et al. \cite{ban2019survey} & & CL, TT & & Hyperledger Burrow, Hyperledger Fabric, Hyperledger Indy, Hyperledger Iroha, Hyperledger Sawtooth \\
BBB \cite{pan2020bbb} & & CL, TT & & Ethereum \\
BBSF \cite{ren2023bbsf} & & CL, TT & FT & Ethereum, Quorum \\
BCadvisor \cite{qian2022bcadvisor} & & CL, TT & & Ethereum, Hyperledger Fabric \\
BLOCKBENCH \cite{dinh2017blockbench} & & CL, TT & FT & Ethereum, Hyperledger Fabric \\
Blockmeter \cite{alom2023blockmeter} & & CL, TT & & Hyperledger Fabric, Hyperledger Sawtooth \\
BlockSim \cite{alharby2020blocksim} & & CL, TT & SBR & Bitcoin, Ethereum \\
C2B2 \cite{kassab2022c2b2} & & CL, TT & & Hyperledger Fabric \\
Caliper \cite{dabbagh2020performance} & & CL, TT & & Hyperledger Fabric \\
COCONUT \cite{geyer2023end} & & CL, TT & & BitShares, Corda Enterprise, Corda Open Source, Hyperledger Fabric, Hyperledger Sawtooth, Quorum \\
Debug-Bench \cite{imran2022research} & & CL, TT & & Ethereum \\
Diablo \cite{gramoli2023diablo} & & CL, TCR, TT & FT & Algorand, Avalanche, Ethereum, Hyperledger Fabric, Quorum, Red Belly, Solana \\
Fu et al. \cite{fu2024quantifying} & TC, CRD, WD & CL, TT & & Algorand, Ethereum \\
GoLevelDB \cite{nakaike2020hyperledger} & & CL, TT & & Hyperledger Fabric \\
Gr{\"a}be et al. \cite{grabe2020not} & & CL$^{\mathrm{a}}$, TT & FT, REL & Ethereum, Hyperledger Indy, Tezos \\
Gromit \cite{nasrulin2022gromit} & & CL, TT & & Algorand, BitShares, Ethereum, Hyperledger Burrow, Hyperledger Fabric, Stellar \\
HyperLedgerLab \cite{chacko2021my} & & CL, TT & & Hyperledger Fabric \\
Jia et al. \cite{jia2022measuring} & BPR, WD & & & Binance Smart Chain, Cardano, Elrond, Fantom, Polygon, Tron \\
Juodis et al. \cite{juodis2024overview} & BPR, TC, CRD, WD & & & Bitcoin, Ethereum \\
Lin et al. \cite{lin2021measuring} & BPR, CRD, WD & & & Bitcoin, Ethereum \\
Prism \cite{liu2021effective} & & CL, TT & & EOS, Ethereum, Hyperledger Burrow, Hyperledger Fabric, Hyperledger Iroha, Hyperledger Sawtooth \\
PTE \cite{meena2019preserving} & & CL, TT & & Hyperledger Fabric \\
Quattrocchi et al. \cite{quattrocchi2024blockchain} & CRD, WD & TT & CoA & Bitcoin, Cardano, Ethereum, Solana \\
SimBlock \cite{nakai2024formulation} & TC, CRD, WD & TT & FT, SBR & Bitcoin \\
SmallBank \cite{sharma2019blurring} & & CL, TT & & Hyperledger Fabric \\
StreamChain \cite{kuhring2021streamchain} & & CL, TT & & Hyperledger Fabric \\
Thakkar et al. \cite{thakkar2018performance} & & CL, TT & FT & Hyperledger Fabric \\
Tsai et al. \cite{tsai2021testbed} & & CL, TT & & Stellar \\
Wu et al. \cite{wu2019information} & BPR & & & Bitcoin, Ethereum \\
\end{tblr}

\vspace{2mm}
\begin{minipage}{\textwidth}
\scriptsize
\textit{Notes.}
BPR = block-proposal randomness (Eq.~\ref{eq:BPR});
CRD = consensus resource distribution (Eq.~\ref{eq:CRD});
TC = token concentration (Eq.~\ref{eq:TC});
WD = wealth distribution (Eq.~\ref{eq:WD_Gini});
CL = confirmation latency (Eq.~\ref{eq:emp});
TCR = transaction confirmation ratio (Eq.~\ref{eq:TCR});
TT = transaction throughput (Eq.~\ref{eq:throughput-emp});
FT = fault tolerance (Eq.~\ref{eq:FT});
REL = reliability (Eq.~\ref{eq:reliability});
SBR = stale block rate (Eq.~\ref{eq:SBR});
CoA = cost of attack (Eq.~\ref{eq:CostofAttack}).
$^{\mathrm{a}}$~Confirmation latency estimated from the required confirmation depth (Eq.~\ref{eq:conf}).
\end{minipage}
\end{table*}

\subsection{Overview of Analysis Approaches and Their Operationalized Constructs}
\label{appdx:ovw-analysis-approaches}

Table~\ref{tab:appendix-constructs} provides an overview of analysis approaches used to investigate blockchain trilemma dimensions, the operationalized constructs they employ, and examples of blockchain systems analyzed using these approaches. The table illustrates how analysis approaches are linked to specific constructs and metrics for quantifying \ac{DoD}, scalability, and security.

Most analysis approaches focus on scalability and commonly operationalize this dimension through \textit{confirmation latency} (e.g., equations~\ref{eq:conf}, \ref{eq:emp}, and \ref{eq:epoch}) and \textit{transaction throughput} (equation~\ref{eq:throughput-emp} and \ref{eq:throughput-cap}). These constructs apply to a broad range of blockchain systems, including permissionless and permissioned systems. Security-related analysis approaches primarily operationalize \textit{fault tolerance} (equation~\ref{eq:FT}), \textit{reliability} (equation~\ref{eq:reliability}), \textit{stale block rate} (equation~\ref{eq:SBR}), and \textit{cost of attack} (equation~\ref{eq:CostofAttack}).

Analysis approaches targeting \ac{DoD} predominantly focus on permissionless blockchain systems, where relevant data are publicly accessible. Commonly operationalized constructs include \textit{block-proposal randomness} (equation~\ref{eq:BPR}), \textit{consensus resource distribution} (equation~\ref{eq:CRD}), \textit{token concentration} (equation~\ref{eq:TC}), and \textit{wealth distribution} (equation~\ref{eq:WD_Gini}). Together, these constructs capture different manifestations of influence and control within blockchain systems.

\thispagestyle{plain}

\end{document}